\documentclass[superscriptaddress,aps,prd,nofootinbib,preprint]{revtex4-1}

\usepackage{graphicx}
\usepackage[colorlinks,allcolors=blue]{hyperref} 
\usepackage{amsmath,amssymb,bm}
\usepackage[T1]{fontenc}
\usepackage{mathtools}

\usepackage[normalem]{ulem}
\usepackage{comment}
\usepackage[dvipsnames]{xcolor}

\begin{document}

\title{QCD Axion Dark Matter from level crossing with refined adiabatic condition}

\author{Kai Murai}
\email{kai.murai.e2@tohoku.ac.jp}
\affiliation{Department of Physics, Tohoku University, Sendai, Miyagi 980-8578, Japan}
\author{Yuma Narita}
\email{yuma.narita.q5@dc.tohoku.ac.jp}
\affiliation{Department of Physics, Tohoku University, Sendai, Miyagi 980-8578, Japan}
\author{Fuminobu Takahashi}
\email{fumi@tohoku.ac.jp}
\affiliation{Department of Physics, Tohoku University, Sendai, Miyagi 980-8578, Japan}
\author{Wen Yin}
\email{wen@tmu.ac.jp}
\affiliation{Department of Physics, Tokyo Metropolitan University, Tokyo 192-0397, Japan}
\affiliation{Department of Physics, Tohoku University, Sendai, Miyagi 980-8578, Japan}

\begin{abstract}
We investigate the level-crossing phenomenon in two-axion systems, where the mass eigenvalues intersect as the mass of one axion increases with the cooling of the universe. This phenomenon can significantly alter the abundance of axions in the early universe. Our study focuses on its impact on the QCD axion and an axion-like particle, identifying viable regions of axion mass and decay constant that explain the observed dark matter. We demonstrate the equivalence of two different bases for describing the axion system in the existing literature. Furthermore, we derive an improved expression for the adiabatic condition that overcomes limitations in earlier formulations. This new formulation is basis-independent, and we numerically validate its effectiveness. Our analysis reveals specific relations between axion masses and axion-photon couplings within the viable region. These relations could potentially serve as a smoking gun signal for this scenario if confirmed experimentally. 
We also find that, using the chiral perturbation model, the thermal friction on the QCD axion might be significantly larger than previously estimated.
{Additionally, we show that a simple model with axion mixing can naturally realize either a heavier or lighter QCD axion.}
 \end{abstract}

\preprint{TU-1252}
\maketitle

\section{Introduction}
Axions are Nambu-Goldstone bosons that arise from the spontaneous breaking of continuous global symmetries. Among them, the QCD axion was introduced to solve the strong CP problem~\cite{Peccei:1977hh,Peccei:1977ur,Weinberg:1977ma,Wilczek:1977pj}. See Refs.~\cite{Kim:2008hd,Arias:2012az,Kawasaki:2013ae,Marsh:2015xka,DiLuzio:2020wdo,OHare:2024nmr} for reviews. Axions are promising dark matter candidates due to their weak interactions and small masses. In string theory, numerous axions appear at low energies, often referred to as axion-like particles (ALPs), especially when coupled to photons~\cite{Witten:1984dg,Conlon:2006tq,Svrcek:2006yi,Arvanitaki:2009fg,Cicoli:2012sz}. In general, these axions exhibit mixing in both kinetic terms and potentials, which has been extensively studied in various contexts such as the level crossing~\cite{Kitajima:2014xla,Daido:2015bva,Daido:2015cba,Ho:2018qur,Murai:2023xjn,Cyncynates:2023esj,Li:2023xkn,Li:2023uvt,Li:2024psa,Li:2024okl}, resonance through non-linear interactions~\cite{Cyncynates:2021xzw}, the misalignment mechanism~\cite{Murai:2023xjn}, decaying dark matter~\cite{Higaki:2014qua}, clockwork mechanism~\cite{Higaki:2015jag,Higaki:2016jjh,Higaki:2016yqk,Long:2018nsl}, inflation~\cite{Choi:2014rja,Higaki:2014pja,Nakayama:2014hga,Daido:2017wwb,Takahashi:2019pqf,Kobayashi:2019eyg,Kobayashi:2020ryx,Narita:2023naj}, and various phenomenological and cosmological aspects~\cite{Chen:2021hfq,Chen:2021jcb}.

To explain the observed dark matter, a right amount of axions must be produced. Among various production mechanisms, the simplest and most natural is the vacuum misalignment mechanism~\cite{Preskill:1982cy,Abbott:1982af,Dine:1982ah}. In the pre-inflationary scenario considered in this paper, the initial values of axion fields are generally displaced from their low-energy potential minima.  When the Hubble parameter becomes comparable to the axion masses, they begin to oscillate around the potential minimum, with the oscillation energy contributing to dark matter. The misalignment mechanism naturally occurs because the presence of an axion in an expanding universe automatically leads to its production.

An important feature of the QCD axion is its temperature-dependent mass. At temperatures above the typical QCD scale, its mass is highly suppressed. As the universe cools down to the QCD scale, the mass of the QCD axion grows, approaching asymptotically to a zero-temperature value. The abundance of the QCD axion produced by the misalignment mechanism depends on the initial misalignment angle, $\theta_\mathrm{i}$, a free parameter in the pre-inflationary scenario with a natural value of order unity. 
The observed dark matter abundance can be naturally explained for the axion decay constant of $f_a \simeq 10^{12}$\,GeV without fine-tuning $\theta_\mathrm{i}$. 
For $f_a$ much larger or smaller than $ 10^{12}$\,GeV, we need to either suppress or enhance the axion abundance. There are several mechanisms, including the stochastic axion scenario~\cite{Graham:2018jyp,Takahashi:2018tdu,Ho:2019ayl,Reig:2021ipa,Murai:2023xjn}, anharmonic effects~\cite{Lyth:1991ub,Kobayashi:2013nva}, non-perturbative production of dark photons~\cite{Agrawal:2017eqm,Kitajima:2017peg}, trapped misalignment mechanism~\cite{Higaki:2016yqk,Nakagawa:2020zjr,DiLuzio:2021gos,Jeong:2022kdr,DiLuzio:2024fyt}, first-order phase transitions generating axion mass~\cite{Nakagawa:2022wwm,Lee:2024oaz}, and level crossing~\cite{Kitajima:2014xla,Daido:2015bva,Daido:2015cba,Ho:2018qur,Murai:2023xjn,Cyncynates:2023esj,Li:2023xkn,Li:2023uvt,Li:2024psa,Li:2024okl}. This paper focuses on the level-crossing phenomenon.

In scenarios with multiple axions, particularly where the QCD axion arises from a combination of the multiple axions, it is plausible for the QCD axion to mix with other axions.
As the QCD axion mass increases as the universe cools, the two mass eigenvalues may intersect. Analogous to the well-known Mikheyev-Smirnov-Wolfenstein effect in solar neutrinos~\cite{Wolfenstein:1977ue,Mikheyev:1985zog}, the two axions could undergo resonant conversion under the adiabatic condition, where the level crossing occurs sufficiently slowly compared to the oscillation time scale.

The possibility of the resonant conversion between axions was first pointed out in Ref.~\cite{Hill:1988bu}, but their estimate of the axion abundance was flawed since they did not realize that the axion number density in the comoving volume is the adiabatic invariant. The axion abundance in this context was first correctly estimated in Ref.~\cite{Kitajima:2014xla}. The adiabatic transition also gives rise to a stochastic dynamics of axions known as axion roulette, where the axion traverses the axion potential peaks and troughs~\cite{Daido:2015bva,Daido:2015cba}. The viable parameter space of the two axions to explain dark matter was studied in detail in Ref.~\cite{Ho:2018qur}, and in particular, it was pointed out that the notion of beat frequency plays a crucial role in the adiabatic condition. 

Recently, the level crossing phenomenon in axion cosmology has attracted more attention and has been studied from various contexts~\cite{Li:2023uvt,Li:2023xkn,Murai:2023xjn,Cyncynates:2023esj,Li:2024psa,Li:2024okl}. In particular, Ref.~\cite{Cyncynates:2023esj} demonstrated that the QCD axion abundance can be enhanced to explain dark matter, making an axion decay constant much smaller than $10^{12}$\,GeV viable. This contrasts with earlier studies~\cite{Kitajima:2014xla,Ho:2018qur}, which explored the possibility of reducing the QCD axion abundance using the resonant conversion. The key distinction between these approaches lies in their implementation of mixing between the QCD axion and another axion.

In this paper, we carefully study the level-crossing phenomenon of the two axions. As a concrete case, we focus on the mixing between the QCD axion and an ALP, and delineate the viable parameter region of the axion masses and decay constants that explains the observed dark matter. In the context of reducing the axion abundance, such a viable region for both the QCD axion and an ALP has already been studied in Ref.~\cite{Ho:2018qur}, but not in the context of increasing the QCD axion abundance. On the other hand, the viable parameter region of the ALP was not given in Ref.~\cite{Cyncynates:2023esj}.
Furthermore, we explicitly show that the two different ways of introducing the mixing of the two axions are in fact equivalent and clarify their relation. Through this analysis, we find that depending on whether one wants to enhance or suppress the axion abundance, one basis is preferred over the other for a natural embedding in the UV theory, since the mixing parameter required for successful resonant conversion can significantly deviate from ${\cal O}(1)$ in an unfavored basis. We also provide a basis-independent adiabatic condition, which turns out to correctly describe the condition under which the resonant conversion takes place. This corrects a somewhat imprecise adiabatic condition used in the literature. 
We also find that the thermal friction on the QCD axion could be larger than previously estimated, due to interactions with pions through axion-pion mixing. This enhanced friction may affect the QCD axion abundance when its decay constant is well below $10^{9}\,$GeV.

The rest of this paper is organized as follows. In Sec.~\ref{sec: review} we briefly review the misalignment mechanism for both the QCD axion and an ALP. In Sec.~\ref{sec: setup} we introduce the mixing between the QCD axion and the ALP, and outline how the level crossing occurs.
In Sec.~\ref{sec: QCD axion DM} we study the dynamics of the axions in detail, using the basis-independent adiabatic condition, and show the viable parameter region.
In Sec.~\ref{sec: basis selection} we confirm that the two ways to introduce the mixing are actually equivalent while the parameter choice and the initial condition become different in the two bases.
In Sec.~\ref{sec: refined adiabatic condition} we compare different adiabatic conditions in the literature and validate the one used in our analysis, which involves the beat frequency.
Sec.~\ref{sec: conclusion} presents our discussion and conclusions.
In the Appendices, we present our numerical methods (Appendix~\ref{sec: numerical calculation}), analyze the thermal friction on the QCD axion (Appendix~\ref{app:friction}), and discuss the implications of axion mixing for heavy QCD axion scenarios (Appendix~\ref{app:mix}).

\section{Review of misalignment mechanism}
\label{sec: review}

Before investigating the dynamics with mixing, we first review the axion abundance in single-axion models. We make two assumptions: first, that the symmetry relevant to the axions is spontaneously broken before or during inflation, resulting in almost homogeneous configurations of the QCD axion field $a$ and the ALP field $\phi$ throughout our universe.\footnote{This assumption simplifies our setup by avoiding contributions from strings and domain walls to the dark matter abundance. Our conclusions about the level crossing remain valid in many cases even when the abundance includes contributions from both topological defects and misalignment. Effects of axion momenta can be straightforwardly incorporated into our analysis.} Second, we assume that all relevant phenomena occur during the radiation-dominated era, where the cosmic temperature $T$ characterizes the temperature of the plasma.

First, we consider the QCD axion abundance due to the misalignment mechanism.
It is evaluated as~\cite{Ballesteros:2016xej}
\begin{align}
    \Omega_a h^2 
    \simeq 
    0.35 \left( \frac{\theta_{a,\mathrm{i}}}{0.001} \right)^{2} \times
    \left\{
    \begin{array}{cc}
    \displaystyle{ \left( \frac{f_a}{3 \times 10^{17}\,\mathrm{GeV}} \right)^{1.17} }
        & \quad(f_a \lesssim 3 \times 10^{17}\,\mathrm{GeV}) 
        \\ 
        \displaystyle{\left( \frac{f_a}{3 \times 10^{17}\,\mathrm{GeV}} \right)^{1.54} }
        & \quad(f_a \gtrsim 3 \times 10^{17}\,\mathrm{GeV})
    \end{array}
    \right.
    \ ,
    \label{eq: a abundance}
\end{align}
where $\Omega_a$ is the current density parameter of the QCD axion, $f_a$ denotes its decay constant, $\theta_{a,\mathrm{i}}$ is the initial value of $\theta_a \equiv a/f_a$,  $h$ is the reduced Hubble constant defined by $h = H_0/(100\,\mathrm{km}\,\mathrm{s}^{-1} \mathrm{Mpc}^{-1})$ with $H_0$ being the Hubble constant, and we have neglected the anharmonic effect. {The non-trivial power dependence of $f_a$ {arises from} the temperature-dependent potential, which will be given in Eq.~(\ref{eq:chi}).}
Thus, for $\theta_{a,\mathrm{i}} = \mathcal{O}(1)$, the observed dark matter abundance, $ \Omega_{\rm DM}h^2 \simeq 0.12$~\cite{Planck:2018vyg}, can be explained by $f_a \simeq 10^{12}$\,GeV.
For larger $f_a$, the initial condition should be tuned to $|\theta_{a,\mathrm{i}}| \ll 1$ to avoid the overabundance.
For smaller $f_a$, dark matter can be accounted for by the QCD axion only if $\theta_{a,\mathrm{i}}$ is tuned as $0< \pi - \theta_{a,\mathrm{i}} \ll 1$, where the anharmonic effect enhances the axion abundance.
However, such a hilltop initial condition suffers from too large isocurvature perturbations~\cite{Lyth:1991ub,Kobayashi:2013nva}.

Next, we consider the abundance of the ALP, $\phi$, due to the misalignment mechanism.
Here, we assume that the potential for $\phi$ is given by
\begin{align}
    V_\phi = m_\phi^2 f_\phi^2 
    \left[ 1 - \cos \left( \frac{\phi}{f_\phi} \right) \right]
    \ ,
\end{align}
where $m_\phi$ and $f_\phi$ are the mass and decay constant of $\phi$, respectively.
Then, the ALP abundance is evaluated as~\cite{ParticleDataGroup:2022pth}
\begin{align}
    \Omega_\phi h^2 
    \simeq 
    0.12  \,\theta_{\phi,\mathrm{i}}^2 \left( \frac{m_\phi}{4.7 \times 10^{-19}\,\mathrm{eV}} \right)^{1/2}
    \left( \frac{f_\phi}{10^{16}\,\mathrm{GeV}} \right)^{2}
    \ ,
    \label{eq: phi abundance}
\end{align}
where $\theta_{\phi,\mathrm{i}}$ is the initial value of $\theta_\phi \equiv \phi/f_\phi$. Here, for simplicity, we have ignored the anharmonic effect as in Eq.~\eqref{eq: a abundance}.

\section{Setup}
\label{sec: setup}

Now we consider the effective Lagrangian for two axions, $a$ and $\phi$, given by 
\begin{align}
    \mathcal{L}
    =
    \frac{1}{2} \partial_\mu a \partial^\mu a
    + \frac{1}{2} \partial_\mu \phi \partial^\mu \phi
    - V(a, \phi)
    \ ,
\end{align}
where the potential is given by 
\begin{align}
    V(a, \phi)
    &=
    V_\mathrm{QCD}(a,\phi) + V_\phi(\phi)
    \nonumber \\
    &=
    \chi(T) \left[ 
        1 - \cos \left( \frac{a}{f_a} + n_\phi \frac{\phi}{f_\phi} \right)
    \right]
    +
    m_\phi^2 f_\phi^2 \left[ 
        1 - \cos \left( \frac{\phi}{f_\phi} \right)    
    \right]
    .
    \label{eq: potential}
\end{align}
Here, we have chosen the origin of $a$ and $\phi$ so that $a = \phi = 0$ becomes the potential minimum.
Since $V_\mathrm{QCD}$ vanishes at $a = \phi = 0$, it corresponds to the strong CP conserving point.
See Appendix~\ref{app:friction} for higher-order corrections to $V_{\rm QCD}$.

The first potential $V_{\rm QCD}$, which mixes $a$ and $\phi$, arises from the coupling to gluons,
\begin{equation}
    \mathcal{L} \supset -\frac{\alpha_s}{8 \pi}\left( \frac{a}{f_a} + n_\phi \frac{\phi}{f_\phi}\right) G \tilde{G},
    \label{eq: axion g coupling}
\end{equation}
where
$\alpha_\text{s}$ is the strong coupling constant, and $G$ and $\tilde{G}$ are the field strength of gluons and its dual, respectively. 
The potential for the ALP $V_{\phi}$
could similarly arise from the coupling to hidden gauge bosons.
Since we set $f_\phi$ so that $V_\phi$ is periodic under $\phi \to \phi + 2\pi f_\phi$, $n_\phi$ is a rational number rather than an integer in general.
If this basis is the natural one in the UV theory, the mixing parameter $n_\phi$ is expected to be of ${\cal O}(1)$.

While we assume that the scale of $V_\phi$ remains constant over the temperature range of interest,
the topological susceptibility of QCD, $\chi(T)$, depends on the temperature as 
\begin{align}
    \chi(T)
    \simeq
    \left\{
        \begin{array}{ll}
            \chi_0 & \quad (T < T_\mathrm{QCD})
            \\
            \chi_0 \left( \frac{T}{T_\mathrm{QCD}} \right)^{-n} & \quad (T \geq T_\mathrm{QCD})
        \end{array}
    \right.
    \ ,
\label{eq:chi}
\end{align} 
where we adopt $\chi_0 = (75.6\,\mathrm{MeV})^4$, $T_\mathrm{QCD} = 153\,\mathrm{MeV}$, and $n = 8.16$~\cite{Borsanyi:2016ksw}. For later convenience, we define the QCD axion mass parameters, $m_a(T)$ and $m_{a,0}$, by
\begin{align}
    m_a(T) 
    \equiv 
    \frac{\sqrt{\chi(T)}}{f_a}
    \ , \quad 
    m_{a,0} 
    \equiv 
    \frac{\sqrt{\chi_0}}{f_a}
    \simeq 
    5.7 \,\mathrm{\mu eV} \left( \frac{f_a}{10^{12}\,\mathrm{GeV}} \right)^{-1}
    \ .
\end{align}

Around  $a = \phi = 0$, the potential can be approximated by the quadratic terms:
\begin{align}
    V(a, \phi)
    &\simeq
    \frac{1}{2}
    \begin{pmatrix}
        a & \phi
    \end{pmatrix}
    \begin{pmatrix}
        m_a^2(T) & \dfrac{n_\phi f_a}{ f_\phi} m_a^2(T)
        \\[1em]
        \dfrac{n_\phi f_a}{f_\phi} m_a^2(T) & m_\phi^2 + \dfrac{n_\phi^2 f_a^2}{f_\phi^2} m_a^2(T)
    \end{pmatrix}
    \begin{pmatrix}
        a \\ \phi
    \end{pmatrix}
    \nonumber \\
    &\equiv 
    \frac{1}{2}
    \begin{pmatrix}
        a & \phi
    \end{pmatrix}
    M^2(T)
    \begin{pmatrix}
        a \\ \phi
    \end{pmatrix}
    \ .
    \label{eq: mass matrix}
\end{align}
We can diagonalize the mass matrix $M^2(T)$ by an orthogonal matrix ${\cal O}$ as
\begin{align}
    \begin{pmatrix}
        m_L^2 & 0
        \\
        0 & m_H^2
    \end{pmatrix}
    =
    \mathcal{O} M^2 \mathcal{O}^T
    \ ,     
\end{align}
where the mass eigenvalues of the heavy and light modes, $m_H(T)$ and $m_L(T)$, are given by
\begin{equation}
    m_{H/L}^2(T) = \frac{m_a^2(T)}{2} \left[1+\frac{1}{r_f^2} + r_m^2 \frac{m_{a,0}^2}{m_a^2(T)} \pm \sqrt{\left(\frac{1}{r_f^2} + r_m^2 \frac{m_{a,0}^2}{m_a^2(T)} - 1 \right)^2 + \frac{4}{r_f^2}} \right]
    \ .
    \label{eq:eigenmass}
\end{equation}
Here, $m_H$ and $m_L$ correspond to the positive and negative sign in the right-hand side, respectively, and we have defined,
\begin{equation}
    r_f \equiv \frac{f_\phi}{n_\phi f_a}, \;\;\;\;\; r_m \equiv \frac{m_\phi}{m_{a,0}},
    \label{eq:rf}
\end{equation}
for convenience. The real orthogonal matrix, $\mathcal{O}$, depends on $T$ and is represented using a mixing angle $\alpha(T)$ as
\begin{align}
    \mathcal{O}(T)
    =
    \begin{pmatrix}
        \cos \alpha & -\sin \alpha
        \\
        \sin \alpha & \phantom{-}\cos \alpha
    \end{pmatrix}
    \ , \quad 
    \tan 2\alpha 
    =
    \frac{2 / r_f}{1 / r_f^2 + r_m^2 [m_{a,0}^2 / m_a^2(T)] - 1}
    \ .
    \label{eq: mixing angle}
\end{align}
The mass eigenstates $a_L$ and $a_H$ corresponding to the mass eigenvalues $m_L$ and $m_H$ are related to $a$ and $\phi$ as
\begin{align}
    \begin{pmatrix}
        a_L \\ a_H
    \end{pmatrix}
    =
    \mathcal{O}
    \begin{pmatrix}
        a \\ \phi
    \end{pmatrix}.
\end{align}
Note that $\alpha$ is not uniquely determined by specifying the value of $\tan 2 \alpha$ alone.
Here, we define $\alpha$ to be a continuous function of $T$ with a boundary condition of $\alpha \to 0$ for $T \to \infty$ (equivalently, as $m_a \to 0$).

\begin{figure}[t!]
    \begin{center}  
        \includegraphics[width=110mm]{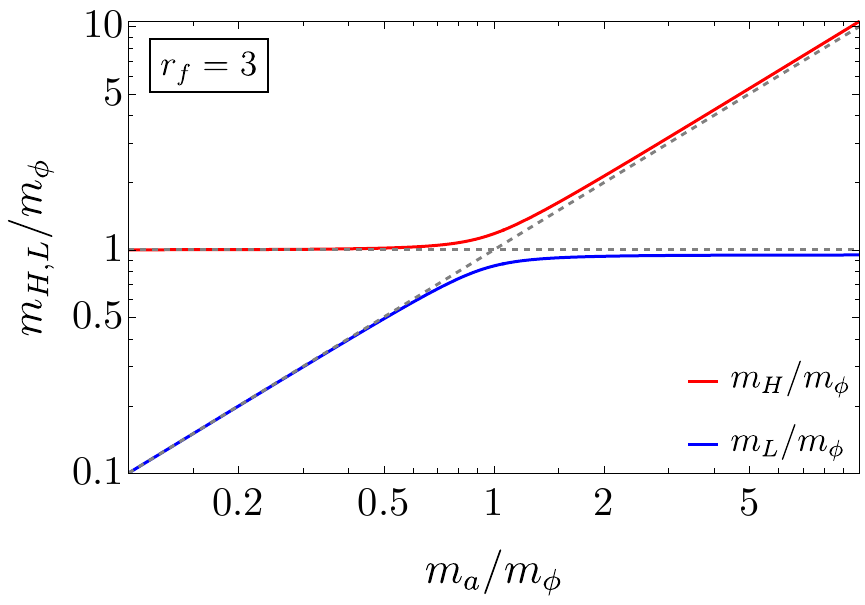}
        \\
        \vspace{5mm}
        \includegraphics[width=110mm]{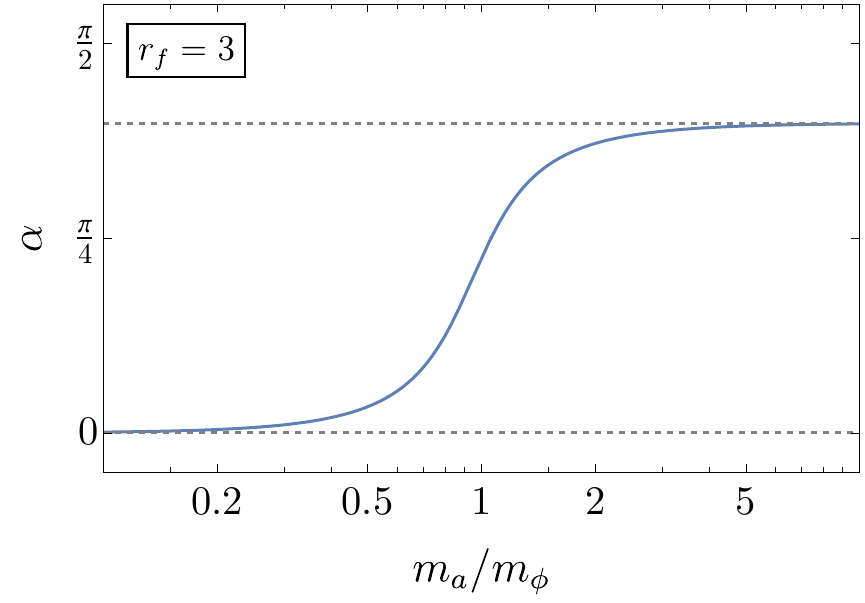}
        \end{center}
    \caption{%
        \textit{Top panel}: the mass eigenvalues as functions of $m_a/m_\phi$ for $r_f = 3$.
        The red and blue lines show the heavier and lighter modes, respectively.
        The gray dashed lines show
        $m_\phi$ and $m_a$.
        \textit{Bottom panel}: the mixing angle $\alpha$ as a function of $m_a/m_\phi$ for $r_f = 3$.
        The horizontal gray dashed lines represent $\alpha = 0$ and $\alpha = \arctan r_f$, respectively.
    }
    \label{fig: level crossing} 
\end{figure}

We show the mass eigenvalues and the mixing angle as a function of $m_a/m_\phi$ in the case of $r_f = 3$ in Fig.~\ref{fig: level crossing}. Here, we treat $m_a/m_\phi$ as a free parameter, while $0 < m_a/m_\phi \leq m_{a,0}/m_\phi$ in our model. When $m_a \gg m_\phi$, $V_\mathrm{QCD}$ determines the mass eigenstates, which are proportional to $\theta_a + n_\phi \theta_\phi$ and the orthogonal direction. On the other hand, when $m_a \ll m_\phi$, $V_\phi$ determines the mass eigenstates, $a$ and $\phi$, with the mass eigenvalues of $m_a$ and $m_\phi$, respectively. When we increase $m_a/m_\phi$ from zero, the level crossing occurs around $m_a \sim m_\phi$. During this transition, the mixing angle evolves from $0$ to $\arctan r_f$ as the mass ratio $m_a/m_\phi$ increases from $0$ to $\infty$. The parameter $r_f$ determines the behavior of the level crossing. As $r_f$ increases, the net rotation angle also increases, approaching $\pi/2$ for $r_f \gg 1$. On the other hand, for $r_f \lesssim 1$, the net rotation angle becomes less than $\pi/2$. In practice, $m_a/m_\phi$ varies over only a finite range, evolving from zero to $m_{a,0}/m_\phi$ as the universe cools. Then, if $m_{a,0} \gtrsim m_\phi$, or equivalently $r_m \lesssim 1$, the mass eigenstates significantly change when $m_a(T) \sim m_\phi$. In the following sections, we discuss such a situation.

Before closing the section, we mention an alternative setup of the potential.
When we choose a basis where the QCD potential depends on one axion, $A$, and the other potential depends on two axions, $A$ and $\Phi$, the potential takes the form
\begin{align}
    \tilde{V}(A, \Phi) 
    =
    \chi(T) \left[ 
        1 - \cos \left( \frac{A}{F_A} \right)    
    \right]
    +
    M_\Phi^2 F_\Phi^2 \left[ 
        1 - \cos \left( N_A \frac{A}{F_A} + \frac{\Phi}{F_\Phi} \right)    
    \right]
    .
    \label{eq: potential another}
\end{align}
This setup has been studied in contexts where the QCD axion abundance is suppressed~\cite{Kitajima:2014xla,Ho:2018qur}. While this appears to be the opposite of Ref.~\cite{Cyncynates:2023esj}, which explores mechanisms to enhance the QCD axion abundance, these two setups are connected via a basis transformation. Although the two bases are mathematically equivalent, the choice of basis naturally reflects the parameter ranges suitable for either suppression or enhancement of the QCD axion abundance. This basis choice often emerges naturally from the corresponding UV-complete theories. When transformed between the two bases, the mixing parameter takes a drastically different value, even though they describe the same physical system. This makes each basis particularly suitable for analyzing its respective scenario. We will discuss this aspect in more detail in Sec.~\ref{sec: basis selection}.

\section{QCD axion dark matter from level crossing}
\label{sec: QCD axion DM}

\subsection{Adiabatic condition}
\label{subsec: adiabatic condition}

We work in the basis of the potential in Eq.~\eqref{eq: potential}, focusing on scenarios where the QCD axion abundance is significantly enhanced through adiabatic level crossing. In what follows, we first establish the conditions necessary for level crossing to occur and then derive the adiabatic condition for level crossing.

\begin{figure}[t!]
    \begin{center}  
        \includegraphics[width=110mm]{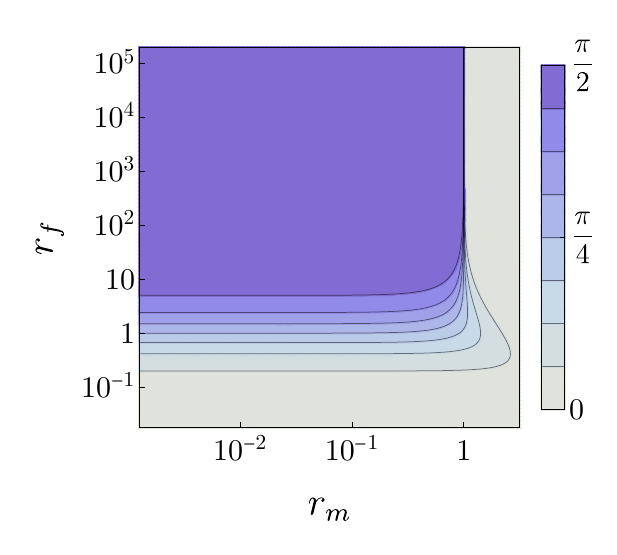}
        \vspace{-7mm}
        \end{center}
    \caption{%
        The contour of $\alpha_0$ in the $(r_m,r_f)$ plane. In the darkest violet region ({left top}), the mass eigenstates undergo a rotation of $\sim \pi/2$, indicating a complete level crossing. As the color lightens, the rotation angle becomes smaller.
    }
    \label{fig: alpha contour} 
\end{figure}

The level crossing induces rotation of eigenstates, and conversely, it does not occur if eigenstates do not rotate sufficiently. 
Therefore, we characterize level crossing by requiring that the extent of eigenstate rotation exceeds a certain threshold:
\begin{equation}
    \Delta \alpha \equiv |\alpha(T \rightarrow 0) - \alpha(T \rightarrow \infty)| > \Delta \alpha_\text{min},
    \label{eq: level crossing condition}
\end{equation}
where $\Delta \alpha_\text{min}$ serves as a threshold for identifying the level crossing. In the basis of \eqref{eq: potential}, we find $\Delta \alpha = \alpha(T \rightarrow 0) \equiv
\alpha_0$.
Since $0 < 2 \alpha(T) < \pi$ in the representation in Eq.~\eqref{eq: mixing angle}, we can rewrite the above condition as 
\begin{equation}
    \cos 2 \alpha_0
    = \frac{1/r_f^2 + r_m^2 - 1}{\sqrt{(1/r_f^2 + r_m^2 - 1)^2 + 4 / r_f^2}} < \cos 2 \Delta \alpha_\text{min}.
    \label{eq: level crossing condition2}
\end{equation}
We show the dependence of $\alpha_0$ on $r_m$ and $r_f$ in Fig.~\ref{fig: alpha contour}.
In particular, if $r_f \gg 1$, the level-crossing condition can be simplified to $r_m \lesssim 1$, which aligns with intuitive expectations. 
Conversely, even if $ r_m \ll 1 $, the level crossing does not occur for $r_f \lesssim 1$, since the eigenstates do not rotate much.
For instance, if we set $\Delta \alpha_\text{min} = \pi / 4$, this condition becomes $r_m < 1$ for $r_f \rightarrow \infty$ and $r_f > 1$ for $r_m \rightarrow 0$.

We define the level-crossing temperature $T_\times$ by
\begin{equation}
    \alpha(T_\times) \equiv \frac{\alpha(T \rightarrow 0) + \alpha (T \rightarrow \infty)}{2}.
    \label{eq: Tlc def}
\end{equation}
From this definition, we obtain
\begin{equation}
    \frac{m_a^2(T_\times)}{m_{a,0}^2} = \frac{r_m^2}{\sqrt{(1 / r_f^2 + r_m^2 - 1)^2 + 4 / r_f^2} + r_m^2} \approx r_m^2 \left( 1 + \frac{1}{r_f^2}\right)^{-1},
    \label{eq: T level crossing}
\end{equation}
where the approximation in the second equality holds for $r_f \gg 1$ and $r_m \ll 1$.

If the field oscillation is rapid enough at the level crossing, the number densities of the heavy and light modes before the level crossing are adiabatically transferred to those after the level crossing, respectively.
To elucidate the condition for the adiabatic level crossing, we compare the timescales of the level crossing and field oscillation. We can estimate the timescale of the level crossing, $\Delta t_\times$, as\footnote{
Note that the time scale of level crossing was estimated using $\alpha$ in Ref.~\cite{Ho:2018qur}. However, the angle appeared through $\cos \alpha$, which is basis-dependent since one can add an arbitrary constant to the angle. Our definition in Eq.~\eqref{eq: time scale lc} is basis-independent.
}
\begin{equation}
    \Delta t_\times = \left| \frac{{\rm d} \alpha}{{\rm d} t}\right|^{-1}_{T = T_\times},
    \label{eq: time scale lc}
\end{equation}
which gives
\begin{align}
    \Delta t_\times 
    &=
    \frac{4 \sqrt{10}}{n \pi} \frac{M_{\rm pl}}{T_\mathrm{QCD}^2} \frac{1}{K(T_\times)}
    F(r_f,r_m)
    \ .
    \label{eq: Delta t cross}
\end{align}
Here, $M_\mathrm{pl} \simeq 2.4 \times 10^{18}$\,GeV is the reduced Planck mass, and $F$ is given by
\begin{align}
    F(r_f,r_m)
    &\equiv 
    \frac{r_m^{4/n}}{r_f}
    \left( \sqrt{(1 - r_m^2 -1/r_f^2 )^2 + 4 / r_f^2} + r_m^2 \right)^{-(2+n)/n}
    \frac{2}{1 + \frac{1 - r_m^2 -1/r_f^2}{\sqrt{(1 - r_m^2 -1/r_f^2 )^2 + 4 / r_f^2}}}
    \nonumber \\
    &\approx 
    \frac{r_m^{4/n}}{r_f},
\end{align}
where the approximation in the second equality holds for $r_f \gg 1$ and $r_m \ll 1$.
The function $K(T)$ is given in terms of relativistic degrees of freedom for energy density and entropy density, $g_{*}(T)$ and $g_{*s}(T)$, as 
\begin{equation}
    K(T) = \frac{\sqrt{g_*(T)} g_{*s}(T)}{3 g_{*s}(T) + T \dfrac{\mathrm{d} g_{*s}(T)}{\mathrm{d} T}}.
    \label{eq: K}
\end{equation}
This function arises from time derivative of temperature, which is determined by the Friedmann equation, 
\begin{equation}
    H^2(T) = \frac{\rho_r(T)}{3 M_{\rm pl}^2} = \frac{1}{3 M_{\rm pl}^2} \frac{\pi^2}{30} g_*(T)  T^4,
    \label{eq: Friedmann eq}
\end{equation}
and the entropy conservation.
Note that the time scale of the level crossing $\Delta t_\times$ is much shorter than the Hubble time when $r_f \gg 1$ and $r_m \ll 1$.

As for the timescale of field oscillations, we consider both the beat period due to the difference in the two mass eigenvalues and the oscillation periods of the eigenstates, following Ref.~\cite{Ho:2018qur}. The adiabatic conversion occurs when $\Delta t_\times$ is sufficiently longer than both the beat period and the oscillation period~\cite{Ho:2018qur},
\begin{equation}
    \Delta t_\times 
    > C_\mathrm{ad} \times
    \max \left[ \frac{2 \pi}{m_L(T_\times)}, \frac{2 \pi}{m_H(T_\times) - m_L(T_\times)}\right]
    \ .
    \label{eq: adiabatic condition}
\end{equation}
Here,  the numerical factor $C_\mathrm{ad}$ of order unity will be determined in Sec.~\ref{sec: refined adiabatic condition}. 
For $r_f \gg 1$ and $r_m \ll 1$, 
we obtain $m_{H/L}(T_\times) \approx m_\phi[1 \pm 1/(2 r_f)]$, and the adiabatic condition is simplified to
\begin{equation}
    \Delta t_\times 
    \gtrsim 
    C_\mathrm{ad} \times
    \frac{2 \pi r_f}{m_\phi}
    \ .
    \label{eq: adiabatic condition approx}
\end{equation}
Thus, the adiabatic condition becomes much tighter than in the case without using the beat frequency.
We will demonstrate why this adiabatic condition \eqref{eq: adiabatic condition} involving the beat frequency is appropriate in Sec.~\ref{sec: refined adiabatic condition}.

Lastly, we comment on a potential effect that breaks the adiabatic conditions due to axion-plasma interactions. The combination $\frac{A}{F_A}=\frac{a}{f_a}+ n_\phi \frac{\phi}{f_\phi}$ interacts with the plasma through the gluon coupling.
In the deconfining phase, the strong sphaleron processes mediate transitions that flip the chiralities of quarks while conserving the baryon number. These processes affect the axial charge density. 
The strong sphaleron effects on the level crossing dynamics can be analyzed by solving the coupled equations of motion for the two axion fields and the Chern-Simons number, taking into account the sphaleron-induced damping effect. However, instead of solving these three coupled equations directly, we can integrate out the Chern-Simons number~\cite{McLerran:1990de,Berghaus:2020ekh} when the sphaleron rate $\sim 3^5\alpha_s^5 T$~\cite{Narita:2023naj} is much faster than
the axion oscillation.
After integrating out the Chern-Simons number, the effective equation of motion for $A$ acquires a dissipative term:
\begin{align}
    \delta \ddot A= -\Gamma_{\rm dis}\dot A .
\end{align}
The dissipation rate is given by~\cite{McLerran:1990de,Berghaus:2020ekh}
\begin{align}\label{dis}
    \Gamma_{\rm dis}\sim \frac{3\alpha_{s}}{4}  \frac{m_f^2}{F_A^2} T,
\end{align} 
where $m_f$ is the {lightest} quark mass below the cosmic temperature $T$~\cite{McLerran:1990de,Berghaus:2020ekh}. 
One might consider using the up quark mass $m_u$ in the formula to estimate the dissipation. However, we argue that this may not be appropriate due to the IR divergence. A proper estimation requires summing over soft gluons. In the framework of chiral perturbation theory, the friction can be estimated as:
\begin{align}
    \label{dischiral}
    \Gamma^{\chi}_{\rm dis}\sim \frac{(100\,{\rm MeV})^2}{F_A^2}T,
\end{align}
when $T\simeq 0.1\,\text{--}\,1$\,GeV (see Appendix~\ref{app:friction}). This expression does not smoothly connect to Eq.~\eqref{dis} when $m_f$ is set to $m_u$. Based on these considerations, we employ the friction form from Eq.~\eqref{dischiral} to estimate the parameter region where dissipation effects become relevant in our discussion.

In the presence of the above dissipation process, we need to see if the dissipation rate is sufficiently small not to affect our estimates of the axion abundance. If the dissipation rate is much smaller than the Hubble parameter, 
\begin{align}
    \Gamma^{ \chi}_{\rm dis}(T)\ll H(T),
    \label{eq: dissipation condition1}
\end{align}    
we can neglect the dissipation effect on the axion abundance. We also impose the adiabatic condition,
\begin{align}
    \Gamma^{\chi}_{\rm dis}(T_\times)\ll m_H(T_\times)-m_L(T_\times ),
\label{eq: dissipation condition2}
\end{align} 
so that the dissipation process does not spoil the adiabatic level crossing. As we will see shortly, this adiabatic condition is met in the region of our interest.

\subsection{Axion abundance}
\label{subsec: axion abundance}

Next, we estimate the axion abundance and the viable parameter region that explains the observed dark matter, assuming adiabatic level crossing. Under this assumption, the comoving number densities of both the heavy and light axions are respectively conserved through the level crossing. In the limit of $r_f \gg 1$ and $r_m \ll 1$, the heavy mode corresponds to the ALP $\phi$ before the level crossing and becomes the QCD axion $a$ after the level crossing. Therefore, the entire number density of the ALP is transferred to the QCD axion through the level crossing. With initial misalignment angles of order unity,  the initial energy density of the ALP tends to be larger than that of the QCD axion, and this process enhances the final QCD axion abundance compared to the case without level crossing.
Consequently, the QCD axion can account for the observed dark matter even with a relatively small decay constant.

Let us consider that both $a$ and $\phi$ have $\mathcal{O}(1)$ initial misalignment angles.
Then, both the heavy and light axions are produced by the misalignment mechanism, and contribute to dark matter. Each axion begins to oscillate at temperature $T_{H/L, \rm osc}$ defined by 
\begin{align}
    3H(T_{H,\rm osc}) &= m_H(T_{H,\rm osc}), \\
    3H(T_{L,\rm osc}) &= m_L(T_{L, \rm osc}),
\end{align}
respectively.
We further assume that both heavy and light modes begin to oscillate sufficiently before the level crossing so that the mass eigenvalues are approximately given by $m_\phi$ and $m_a(T_{L, \rm osc})$. 
In most of the region of interest, this approximation does not change the timing by more than 10\%. We note that, if $H(T_\times) \gtrsim m_L(T_\times)$ the adiabatic level crossing cannot happen since the light mode does not oscillate until the level crossing. Then, the energy densities of the heavy and light modes at the onset of oscillation are estimated by
\begin{align}
    \rho_H(T_{H, {\rm osc}}) &\simeq \frac{1}{2} m_\phi^2 f_\phi^2 \theta_{\phi, {\rm i}}^2, 
    \label{eq: heavy osc} \\
    \rho_L(T_{L, {\rm osc}}) &\simeq \frac{1}{2} m_a^2(T_{L,\rm osc}) f_a^2 \theta_{a, {\rm i}}^2,
    \label{eq: light osc}
\end{align}
respectively.
We define the axion yields of the mass eigenstates by
\begin{align}
    Y_H(T) = \frac{\rho_H(T)}{m_H(T) s(T)}, \;\;\;
    Y_L(T) = \frac{\rho_L(T)}{m_L(T) s(T)},
\end{align}
where $s(T) = {2 \pi^2 g_{*s}(T) T^3}/{45}$ is the entropy density. 
These quantities are separately conserved after the onset of oscillations as long as the level crossing is adiabatic. Then, we can estimate the present density parameters of both heavy and light axions as
\begin{align}
    \Omega_H 
    &= \frac{m_{H,0} Y_H(T_{H,{\rm osc}}) s_0 }{\rho_{\rm crit}} = 
    \frac{\sqrt{\chi_0} m_{H,0} f_a}{2 \rho_{\rm crit}}  
    \frac{s_0}{s(T_{H,{\rm osc}})} 
    n_\phi^2 r_m r_f^2 \theta^2_{\phi,{\rm i}},
    \label{eq: heavy abundance} 
    \\
    \Omega_L 
    &=\frac{m_{L,0} Y(T_{L,{\rm osc}}) s_0 }{\rho_{\rm crit}} = 
    \frac{\sqrt{\chi_0} m_{L,0}  f_a}{2 \rho_{\rm crit}} 
    \frac{s_0}{s(T_{L,{\rm osc}})}
    \frac{m_a(T_{{\rm osc}, L})}{m_{a,0}}
    \theta^2_{a,{\rm i}} ,
    \label{eq: light abundance}
\end{align}
where $m_{H,0}$ and $m_{L,0}$ are the mass eigenvalues at the zero temperature,  $\rho_{\rm crit}$ is the present critical density, and $s_0$ is the present entropy density.
The total dark matter abundance is given by the sum of their contributions, $\Omega_{\rm DM} = \Omega_H+ \Omega_L$. 
When $r_f \gg 1$, 
the heavy axion abundance is significantly enhanced compared to the light axion, and
we obtain
\begin{equation}
    \label{eq: DM abundance}
    \Omega_{\rm DM} h^2 \;{\simeq}\; 0.12  \ n_\phi^2 \theta_{\phi,{\rm i}}^2 \left( \frac{r_m}{0.1} \right)^{-\frac{1}{2}} \left( \frac{r_f}{100} \right)^{2} \left( \frac{f_a}{10^{10}\,{\rm GeV}} \right)^{\frac{3}{2}},
\end{equation}
where we have used $g_*(T_{H,{\rm osc}}) = g_{*s}(T_{H,{\rm osc}}) = 60$, and $\rho_{\rm crit} / s_0 \simeq 3.644 \times 10^{-9}\,\text{GeV} \ h^2$.

\begin{figure}[t!]
    \begin{center}  
        \vspace{5mm}
        \includegraphics[width = 81mm]{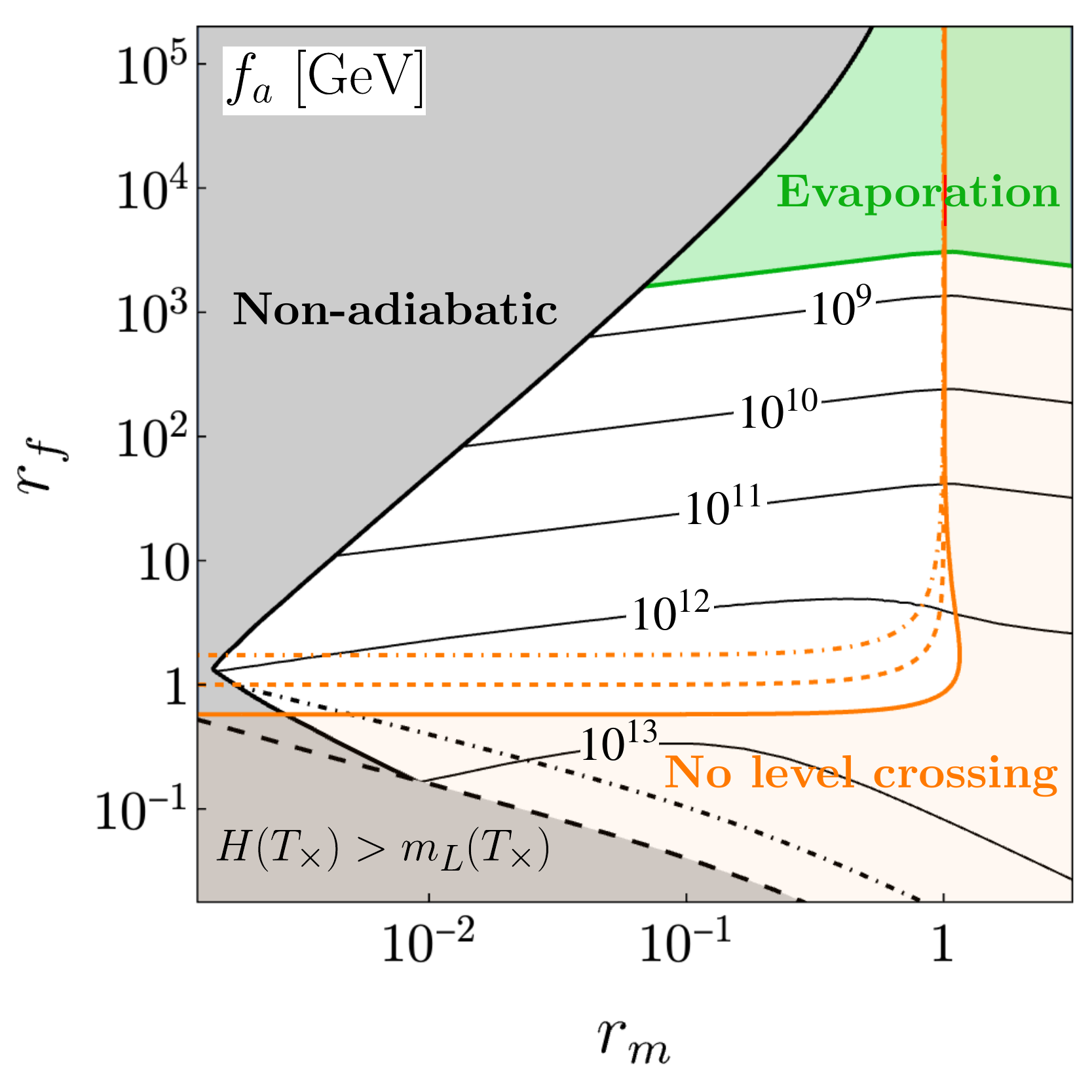}
        \includegraphics[width = 81mm]{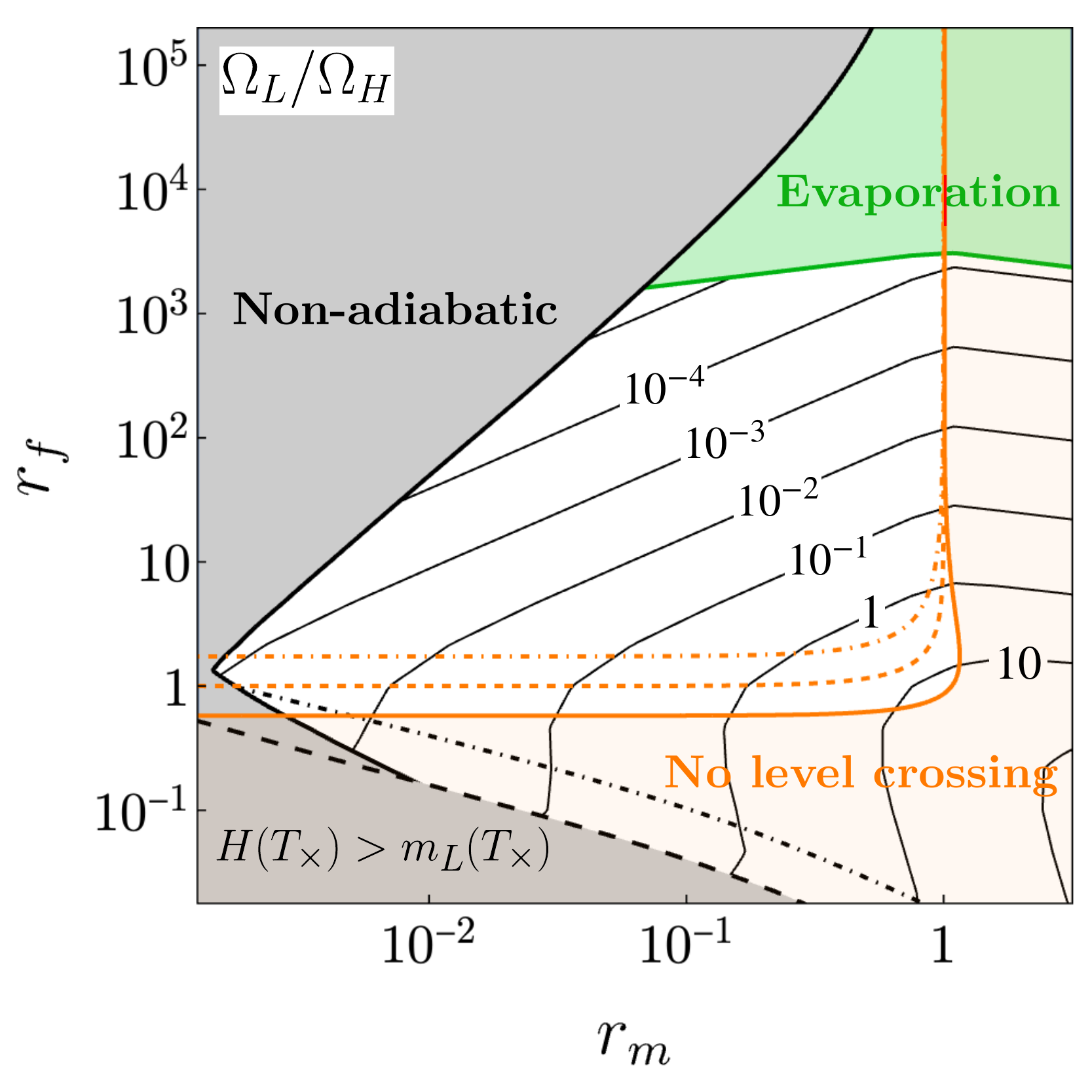}
    \end{center}
    \caption{The left panel represents the contour of $f_a$ in the unit of $\mathrm{GeV}$ to explain all dark matter abundance with respect to $r_m$ and $r_f$, while the right panel represents the contour of the ratio, $\Omega_L/\Omega_H$.
    In the gray region, the condition~\eqref{eq: adiabatic condition} is not satisfied for the corresponding value of $f_a$, and the level crossing is non-adiabatic. 
    Below the black dot-dashed line, $m_L$ differs from $m_a$ by more than 10\% when the light mode begins to oscillate, and the analysis here uses $m_L$ for accurate results.
    In the orange region, the eigenstates do not rotate sufficiently, and the level crossing does not occur.
    We set the threshold as $\Delta \alpha_\text{min} = \pi/6$, $\pi/4$, and $\pi/3$ for the orange solid, dashed, and dot-dashed lines, respectively.
    In the green region, the axion abundance is subject to the evaporation.}
    \label{fig: fa contour} 
\end{figure}

Finally, we derive the viable parameter space for explaining the observed dark matter via adiabatic level crossing.
The abundance estimates are given in Eqs.~\eqref{eq: heavy abundance} and \eqref{eq: light abundance}, which are valid as long as the adiabaticity condition in Eq.~\eqref{eq: adiabatic condition} and $H(T_\times)  < m_L(T_\times)$ are satisfied. 
We note that the estimations of the heavy and light axion abundances remain valid even when level crossing does not occur according to our criterion, provided that the adiabaticity condition is satisfied.

Fig.~\ref{fig: fa contour} shows the contours of $f_a$ (left panel) and $\Omega_L/\Omega_H$ (right panel) that explain the observed dark matter abundance in the $(r_m, r_f)$ plane, where we estimate the axion abundances without the approximation of axion masses
for the initial field amplitudes  $a_{H,\text{i}} = f_\phi$ and $a_{L,\text{i}} = f_a$.
The white region in the figure is 
 the physically viable parameter space where
the adiabaticity condition is satisfied and level crossing occurs.
In the orange region, the level crossing does not occur in our criterion. 
In the gray region, the level crossing is non-adiabatic, and our estimate for the axion abundance becomes invalid. We have used the numerical coefficient {$C_{\rm ad}$} which is numerically determined in Sec.~\ref{sec: refined adiabatic condition}. Below the black dot-dashed line, the light axion mass deviates from the QCD axion mass by more than 10\% when the light mode begins to oscillate. Consequently, the above analytical estimates become invalid.
In the gray region with $r_f \gg 1$ and $r_m \ll 1$, the level-crossing time scale $\Delta t_\times$ from Eq.~\eqref{eq: time scale lc} is so short that the axions do not have sufficient time to oscillate. On the other hand, in the gray region with $r_f \lesssim 1$ and $r_m \ll 1$, the level crossing lasts longer. However, it occurs much before the QCD axion mass intersects the ALP mass, $m_a(T) = m_\phi$. Then, the QCD axion mass is too small at the level crossing for the light axion to oscillate repeatedly, and thus the level crossing is non-adiabatic.
The inequality in Eq.~\eqref{eq: dissipation condition1} is violated in 
the green region, where {the thermal effect}
would dissipate the energy of heavy axion condensate since the heavy axion after the level crossing is {(relatively) }strongly coupled to QCD.
As a result, $Y_H$ is not conserved, and our estimate of the axion abundance {may be} invalid in the green region.%
\footnote{The adiabatic condition in Eq.~\eqref{eq: dissipation condition2}
is automatically satisfied in the white region in Fig.~\ref{fig: fa contour}. This is because, in the region where adiabatic phenomenon occurs, the beat frequency tends to be relatively high, ensuring that this condition holds.}
To evaluate the dark matter abundance within these regions,
{more careful estimations} 
will be required. 
In a large part of the white region, the heavy mode gives a dominant contribution to dark matter, and thus the contours of $f_a$ can be obtained using Eq.~\eqref{eq: DM abundance}. In contrast, the light axion may contribute to the dark matter abundance for $r_f = \mathcal{O} (1)$.

\subsection{Axion-photon coupling}
\label{subsec: axion-photon coupling}

\begin{figure}[t!]
    \begin{center}  
        \includegraphics[width = 81mm]
        {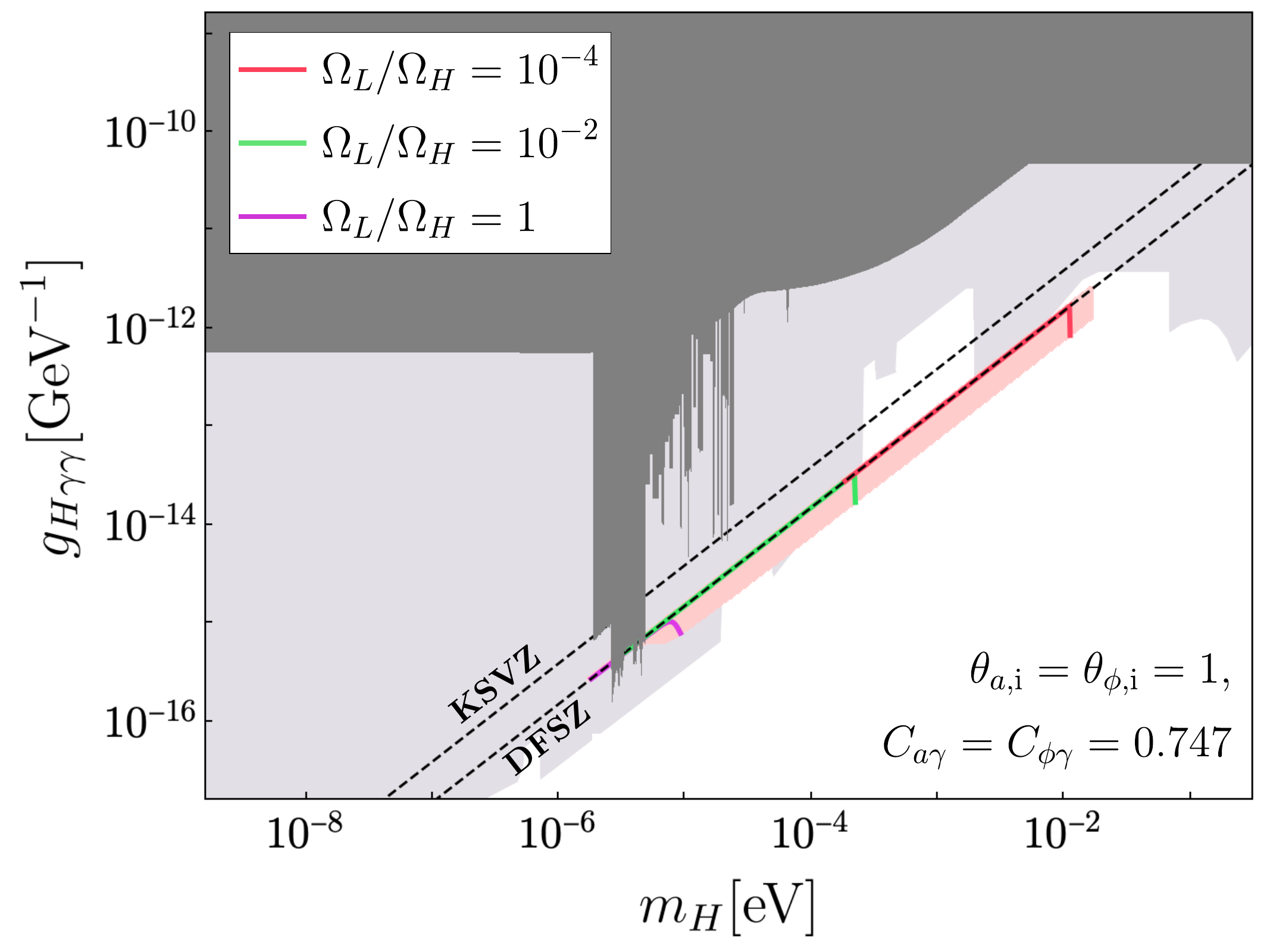}
        \includegraphics[width = 81mm]
        {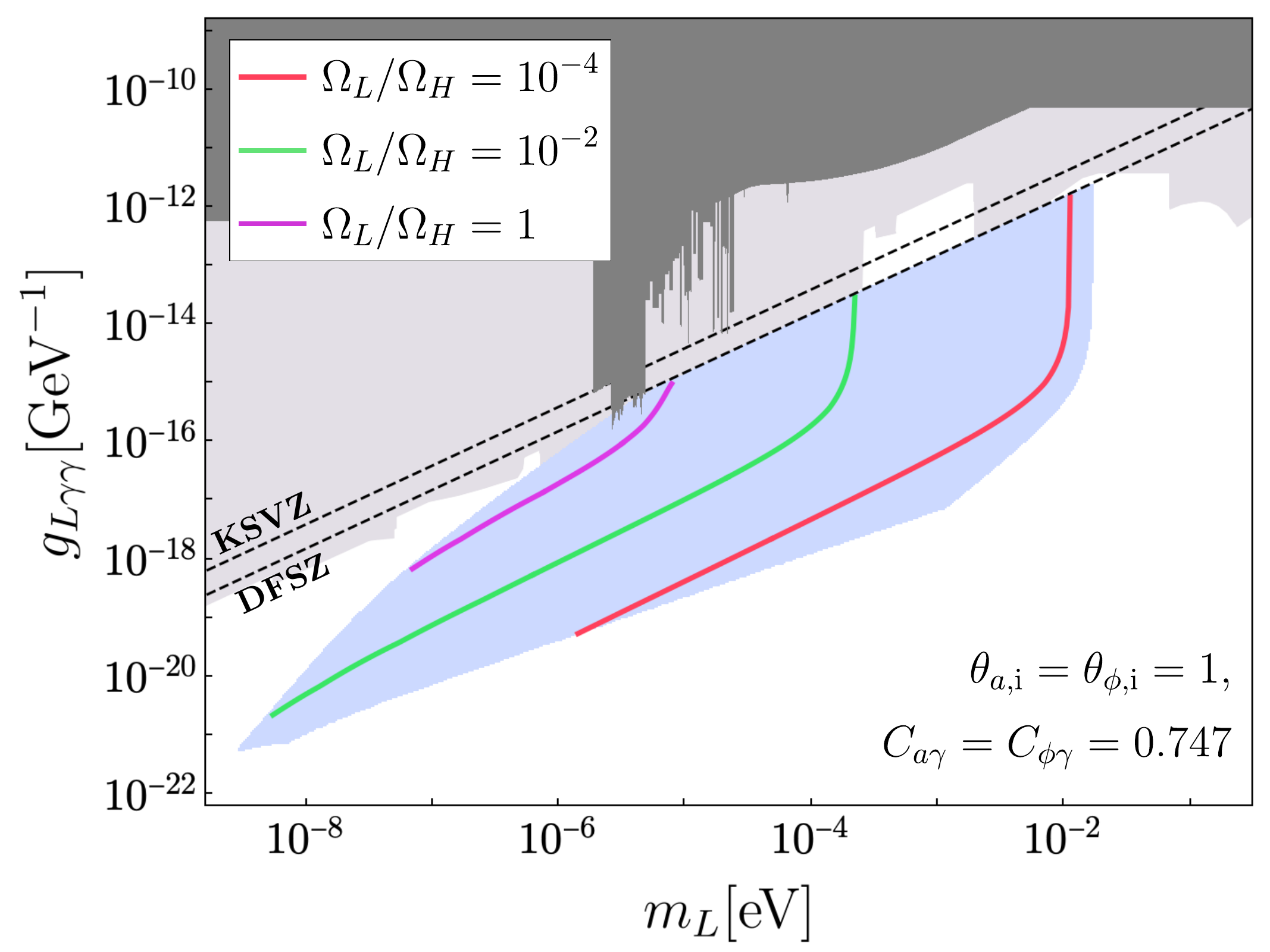}
    \end{center}
    \caption{%
    Mass and photon coupling of the heavy and light modes.
    These panels show the case of $C_{a \gamma} = C_{\phi \gamma} = 0.747$ which corresponds to the DFSZ model~\cite{Dine:1981rt, Zhitnitsky:1980tq}.
    In the red and blue regions, the axions can explain all dark matter via the adiabatic level crossing in the white region in Fig.~\ref{fig: fa contour}. 
    The colored lines correspond to fixed values of $\Omega_L/\Omega_H$.
    The black dashed lines show the QCD axion-photon coupling in the KSVZ model~\cite{Kim:1979if, Shifman:1979if} and the DFSZ model. 
    The dark-gray regions represent constraints from current observations and experiments, while the light-gray transparent regions indicate the projected sensitivity of upcoming experiments. These regions are taken from the data compiled in Ref.~\cite{AxionLimits}. Note that the bounds and future prospects are shown assuming each mode (heavy or light) composes all dark matter. For subdominant components, these bounds should be rescaled accordingly. 
    }
    \label{fig: reliable region} 
\end{figure}

\begin{figure}[t!]
    \begin{center}  
        \vspace{5mm}
        \includegraphics[width = 81mm]
        {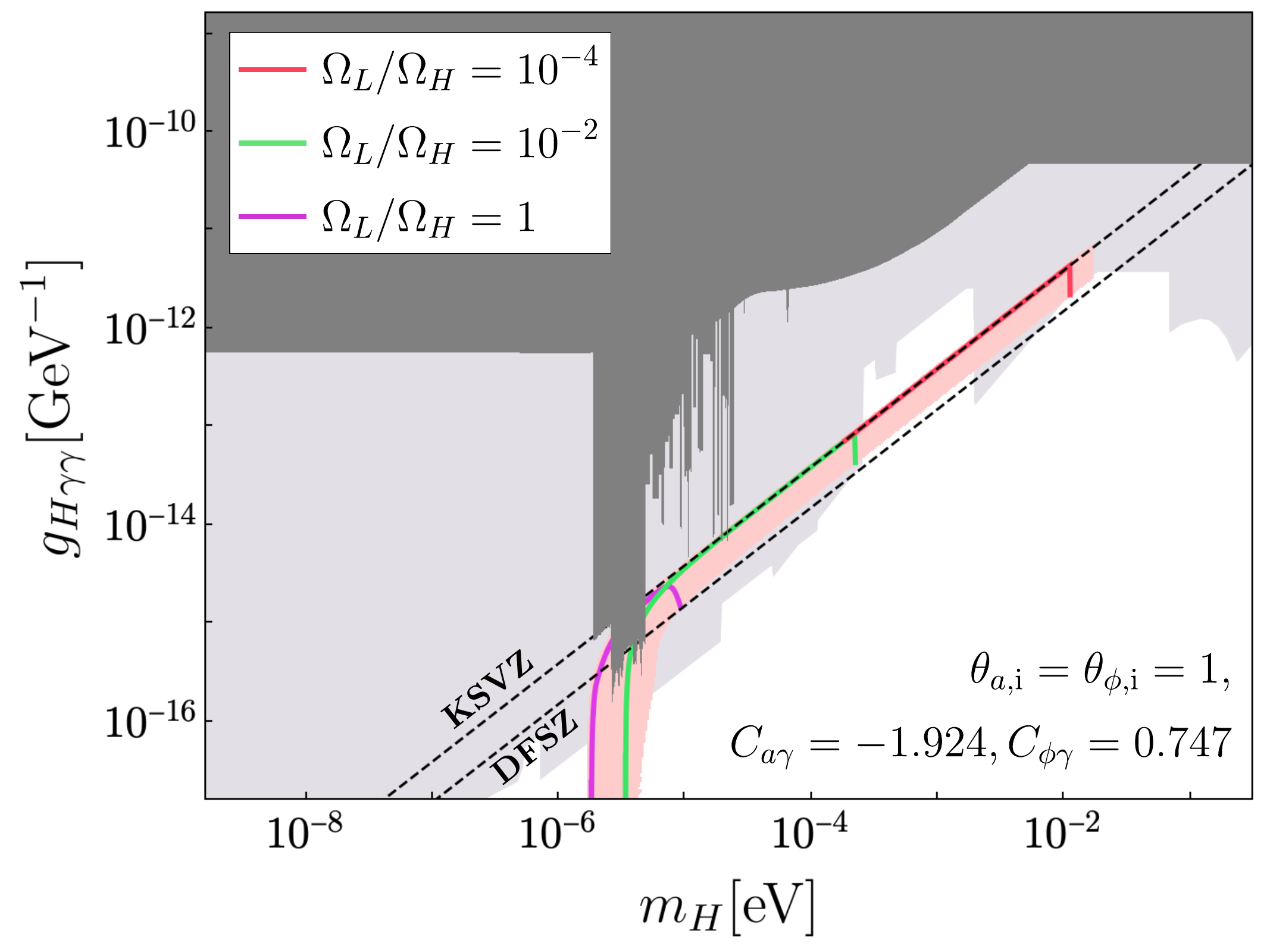}
        \includegraphics[width = 81mm]
        {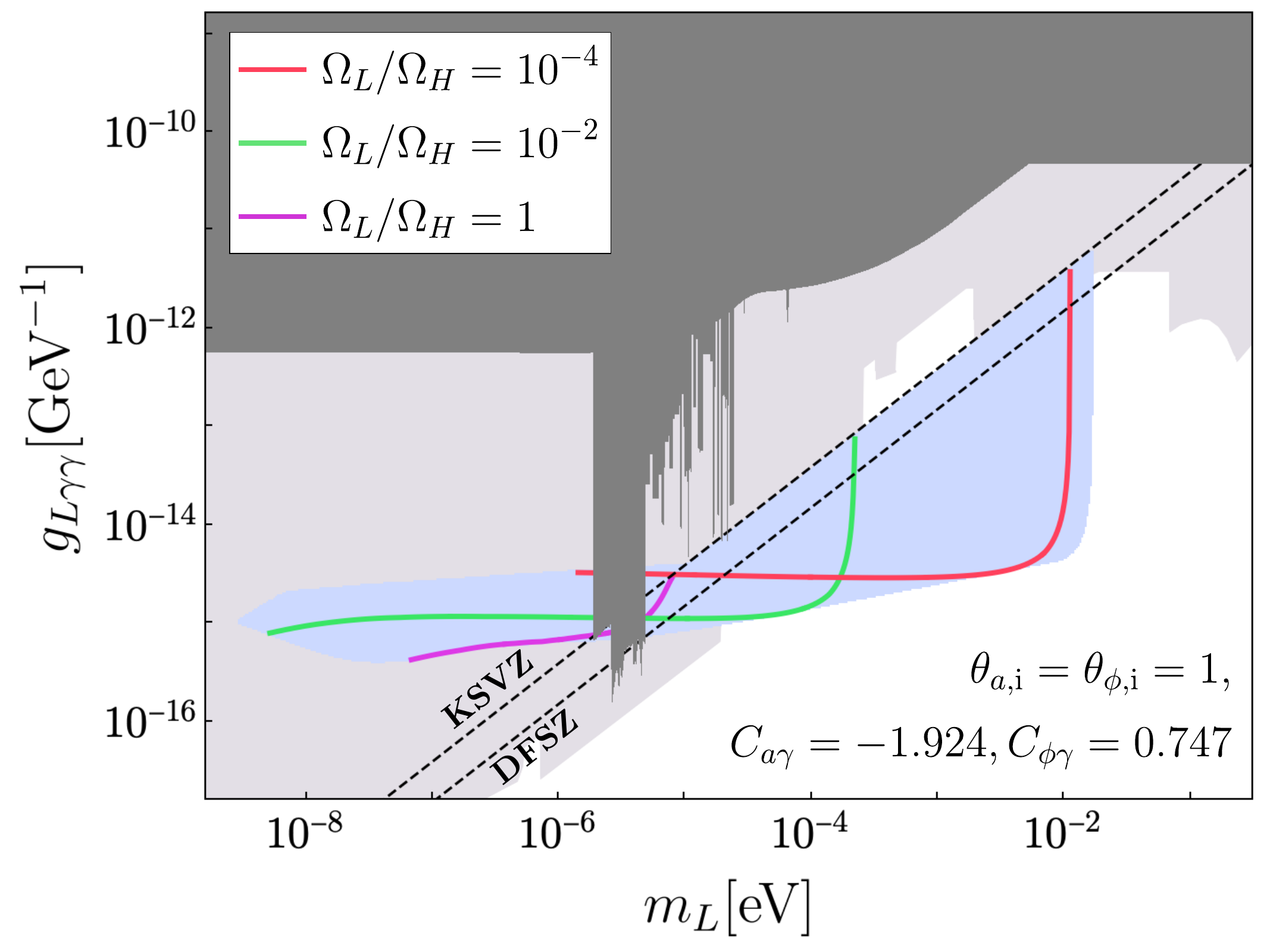}
    \end{center}
    \caption{%
    The same figures as Fig.~\ref{fig: reliable region}, but for $C_{a \gamma} = -1.92, \ C_{\phi \gamma} = 0.747.$
    }
    \label{fig: reliable region2} 
\end{figure}

Additionally, we discuss the verifiability of the axions in our scenario.
Various axion search experiments rely on the axion-photon coupling.
The Lagrangian describing the interaction between axions and photons (at the renormalization scale below the confinement scale) is given by 
\begin{equation}
\begin{split}
    \mathcal{L}_{a\-\gamma\-\gamma} &= -\frac{\alpha_{\rm em}}{8 \pi} \left( C_{a \gamma} \frac{a}{f_a} + C_{\phi \gamma} n_\phi \frac{\phi}{f_\phi} \right) F_{\mu \nu} \tilde{F}^{\mu \nu},\\
    &\equiv -\frac{1}{4}(g_{L\gamma\gamma} a_L + g_{H \gamma \gamma} a_H) F_{\mu \nu} \tilde{F}^{\mu \nu},
\end{split}
\end{equation}
where $g_{H \gamma \gamma}$ and $g_{L \gamma \gamma}$ are the coupling constants of the heavy and light axions,
\begin{align}
    \label{eq: H photon coupling}
    g_{H \gamma \gamma} &= \frac{\alpha_{\rm em}}{2 \pi f_a} \left(C_{a \gamma} \sin{\alpha_0} + C_{\phi \gamma} \frac{\cos{\alpha_0}}{r_f}\right), \\
    \label{eq: L photon coupling}
    g_{L \gamma \gamma} &= \frac{\alpha_{\rm em}}{2 \pi f_a} \left(C_{a \gamma} \cos{\alpha_0} - C_{\phi \gamma} \frac{\sin{\alpha_0}}{r_f}\right).
\end{align}
In our study, we consider two cases:
$ C_{a \gamma} = \ C_{\phi \gamma} = 0.747\, (\simeq 8/3 - 1.92) $, and $C_{a \gamma} = - 1.92, \ C_{\phi \gamma} = 0.747$. The first case represents the configuration derived from the DFSZ model ~\cite{Dine:1981rt, Zhitnitsky:1980tq}, where the axion-photon coupling takes a specific form related to the axion-gluon coupling \eqref{eq: axion g coupling} due to anomaly matching in the GUT symmetric phase. The second case serves as an example that does not impose the GUT configuration.
Figures~\ref{fig: reliable region} and \ref{fig: reliable region2} show the viable regions for heavy and light axions with respect to their mass and photon coupling in each case. We also include lines corresponding to fixed values of $\Omega_L / \Omega_H$. 
{At the vertical segments of these lines on the heavy side, the heavy axion mass is comparable to the light axion mass, despite the hierarchy of the decay constants. 
This is because when the {axion masses become} comparable, the system has an ${\cal O}(1)$ mixing {that reduces} the heavy axion coupling slightly and {enhances} the light axion to the level of the heavy axion.}
For $ m_H > 10^{-5} \, \rm eV $, the heavy axion follows the QCD axion band, where $ |g_{H \gamma \gamma }| = \alpha_{\mathrm{em}} / (2 \pi f_a) $ is independent of $ f_\phi $ in both Figs.~\ref{fig: reliable region} and \ref{fig: reliable region2}. This occurs because most of dark matter is dominated by the heavy axion which can be regarded as QCD axion and its abundance must be enhanced via the level crossing to explain the observed dark matter for $m_H > 10^{-5} \ \text{eV}$. Such a situation is realized with $r_f \gg 1$ and $r_m < 1$ (see Fig.~\ref{fig: fa contour}). The heavy axion in this scenario can be probed by various future axion search experiments, such as BREAD \cite{BREAD:2021tpx} and MADMAX \cite{Caldwell:2016dcw}.

As $ r_f $ approaches unity, the final heavy mode direction deviates from the QCD axion.
Then, the photon coupling with the light axion tends to be canceled if $C_{a \gamma} = C_{\phi \gamma}$.
On the other hand, if $C_{a \gamma} \ne C_{\phi \gamma}$ as in the second example, the light axion tends to have a stronger coupling to photons than the heavy axion especially for relatively large ratio, $\Omega_L / \Omega_H$ (see the right panel of Fig.~\ref{fig: reliable region2}).
This is because, in this case, the light and heavy axions are mainly given by $a$ and $\phi$, respectively, and $m_L$ extends to light mass regions due to the mixing effect. 
In this case, the light axion is {also} detectable in future experiments 
such as DM Radio~\cite{DMRadio:2022pkf} {even with the suppressed abundance by noting that the sensitivity reach scales with $g^{\rm reach}_{L \gamma\gamma} \propto 1/\sqrt{\Omega_L}$ (see also Appendix.~\ref{app:mix} the impact on the other couplings via mixing and the identification with the heavy/light QCD axion).}

In the right panel of Fig.~\ref{fig: reliable region2}, we can see that the $g_{L \gamma\gamma}$ becomes almost constant when we fix $\Omega_L/\Omega_H$. This can be understood as follows. With the resonant conversion, we have $\Omega_L \sim \Omega_a m_\phi/m_{a,0} \propto m_\phi f_a f_a^{1.17}, \Omega_H \sim \Omega_\phi m_{a,0}/m_\phi \propto f_\phi^2 m_\phi^{-1/2} f_a^{{-1}} $. By requiring $\Omega_L/\Omega_H=$const and $\Omega_H+\Omega_L=$const and approximating $1.17\sim 1$, we get the scaling, $f_\phi\sim $ const, leading to $g_{L\gamma\gamma}\sim $ const, and $m_\phi\propto f_a^{-2}$. {Note that we have used the fact that $g_{L \gamma \gamma}$ is dominated by the second term in Eq.~(\ref{eq: L photon coupling}) in this case. }

\section{Basis selection and its cosmological implication}
\label{sec: basis selection}

In the previous section, we mainly examined how the QCD axion abundance can be enhanced using the potential given in Eq.~\eqref{eq: potential}.
An alternative scenario, where the QCD axion abundance is suppressed was studied based on the potential~\eqref{eq: potential another} in Refs.~\cite{Kitajima:2014xla,Ho:2018qur}.
Here, we discuss the equivalence of these two apparently different setups and show that natural model parameters can achieve either enhancement or suppression of the axion abundance in only one of these two frameworks.

For the readability, let us write the two potentials again;
\begin{align}
      V(a, \phi)
    &=
    \chi(T) \left[ 
        1 - \cos \left( \frac{a}{f_a} + n_\phi \frac{\phi}{f_\phi} \right)
    \right]
    +
    m_\phi^2 f_\phi^2 \left[ 
        1 - \cos \left( \frac{\phi}{f_\phi} \right)    
    \right],\\
      \tilde{V}(A, \Phi) 
    &=
    \chi(T) \left[ 
        1 - \cos \left( \frac{A}{F_A} \right)    
    \right]
    +
    M_\Phi^2 F_\Phi^2 \left[ 
        1 - \cos \left( N_A \frac{A}{F_A} + \frac{\Phi}{F_\Phi} \right)    
    \right]
    .
\end{align}
We first show the equivalence between these bases. 
These two bases are related by the rotation matrix $\mathcal{O}_F$ as 
\begin{align}
    \begin{pmatrix}
        a \\ \phi
    \end{pmatrix}
    =
    \begin{pmatrix}
        \cos \alpha_F & -\sin \alpha_F
        \\
        \sin \alpha_F & \phantom{-}\cos \alpha_F
    \end{pmatrix}
    \begin{pmatrix}
        A \\ \Phi
    \end{pmatrix}
    \equiv
    \mathcal{O}_F
    \begin{pmatrix}
        A \\ \Phi
    \end{pmatrix}
    ,
\end{align}
where the rotation angle $\alpha_F$ is given by 
\begin{equation}
    \tan{\alpha_F} = \frac{N_A F_\Phi}{F_A}.
\end{equation}
Here, we set the range of $\alpha_F$ to be $-\pi/2 < \alpha_F < \pi/2$.
The parameters are related as
\begin{equation}
\begin{gathered}
    f_a 
    =
    \sqrt{F_A^2 + N_A^2 F_\Phi^2}
    =
    \frac{F_A}{\cos \alpha_F}
    \ , \quad 
    f_\phi 
    = 
    \frac{F_A F_\Phi}{\sqrt{F_A^2 + N_A^2 F_\Phi^2}}
    =
    F_\Phi \cos \alpha_F
    \ , 
    \\
    m_\phi 
    =
    \frac{\sqrt{F_A^2 + N_A^2 F_\Phi^2}}{F_A} M_\Phi
    =
    \frac{M_\Phi}{\cos \alpha_F}
    \ , \quad 
    n_\phi
    =
    \frac{N_A F_\Phi^2}{F_A^2 + N_A^2 F_\Phi^2}
    =
    \frac{\sin^2 \alpha_F}{N_A}
    \ .
\end{gathered}
\label{eq: f to F}
\end{equation}
We define $R_F$ and $R_M$ similarly to  $r_f$ and $r_m$ as 
\begin{equation}
    R_F  \equiv  \frac{N_A F_\Phi}{F_A}, \;\;\;\;\;  R_M  \equiv \frac{M_\Phi}{M_A},
    \label{eq:RF}
\end{equation}
where we have defined $M_A = \sqrt{\chi_0}/F_A$. They are related to each other as
\begin{align}
    r_f 
    &=
    \frac{f_\phi}{n_\phi f_a}
    =
    \frac{F_A}{N_A F_\Phi}
  =
    R_F^{-1}
    \ ,
    \\
    r_m
    &=
    \frac{m_\phi f_a}{\sqrt{\chi_0}}
    =
    \left(1 + \frac{N_A^2 F_\Phi^2}{F_A^2}  \right)
    \frac{F_A M_\Phi}{\sqrt{\chi_0}}
  =
    \left(1 + R_F^2  \right) R_M
    \ .
\end{align}
Thus, the condition for the level crossing, $r_f \gg 1$ and $r_m \ll 1$ in the basis of Eq.~\eqref{eq: potential}
correspond to $R_F \ll 1$ and $R_M \ll 1$ in the basis of Eq.~\eqref{eq: potential another}.
Using these relations, we can study the dynamics of the model with the potential of Eq.~\eqref{eq: potential another} by using the potential \eqref{eq: potential} and vice versa.
In fact, when we consider $r_f \gg 1$, we obtain $\cos \alpha_F \simeq 1$, and thus the mass parameters and decay constants take similar values in both basis.
Note, however, that  $N_A$ becomes much smaller than unity for $n_\phi = \mathcal{O}(1)$. 
In other words, the setup in the previous section where the QCD axion abundance is enhanced by the level crossing corresponds to a setup with an unnaturally small $N_A$ in the $(A,\Phi)$ basis. This is the reason why the enhancement of the QCD axion abundance was not seen for the natural choice of the mixing parameter in Refs.~\cite{Kitajima:2014xla,Ho:2018qur}.

Let us focus on the estimation of dark matter abundance in the two bases. Since the conditions for level crossing given in Eqs.~\eqref{eq: level crossing condition}, \eqref{eq: Tlc def}, \eqref{eq: time scale lc}, and \eqref{eq: adiabatic condition} are basis-independent (see also Sec.~\ref{sec: refined adiabatic condition}), we can analyze these phenomena in any chosen basis. However, care must be taken when estimating dark matter abundance, as both the initial phases and the abundance of each axion depend on the choice of basis.
The relation of the axion angles in the two bases is given by 
\begin{align}
    \Theta_A 
    &\equiv 
    \frac{A}{F_A}
    =
    \frac{a \cos \alpha_F + \phi \sin \alpha_F}{f_a \cos \alpha_F}
    =
    \theta_a + n_\phi \theta_\phi 
    \ ,
    \\
    \Theta_\Phi
    &\equiv 
    \frac{\Phi}{F_\Phi}
    =
    \frac{- a \sin \alpha_F + \phi \cos \alpha_F}{f_\phi/\cos \alpha_F}
    =
    - \frac{1}{n_\phi(1+r_f^2)} \theta_a 
    + \frac{r_f^2}{1 + r_f^2} \theta_\phi
    \ ,
\end{align}
or equivalently,
\begin{align}
    \theta_a &=\frac{1}{1+R_f^2}\Theta_A-\frac{R_F^2}{N_A(1+R_F^2)}\Theta_\Phi,
    \\
    \theta_\phi
    &=
    N_A \Theta_A + \Theta_\Phi.
\end{align}
If $r_f \gg 1$ and $n_\phi = \mathcal{O}(1)$, these relations reduce to $\Theta_A \simeq \theta_a + n_\phi \theta_\phi$ and $\Theta_\Phi \simeq \theta_\phi$.

To be concrete, we consider the description of the phenomena discussed in Sec.~\ref{subsec: axion abundance} based on Eq.~\eqref{eq: potential another}.
Now, the abundance of the heavy axion significantly exceeds that of the light axion due to the hierarchy of the decay constants ($r_f \gg 1$) in Eq.~\eqref{eq: heavy abundance}. Consequently, the QCD axion constitutes the primary component of dark matter even if its mass is heavier than $\mathcal{O}(10^{-6}) \, \text{eV}$, in contrast to 
the standard misalignment mechanism. For $r_f \gg 1$, we express the abundances of both axions in the basis of Eq.~\eqref{eq: potential another} as follows:
\begin{align}
    \Omega_H 
    &\approx
    \frac{\chi_0}{2 \rho_{\rm crit}}  
    \frac{s_0}{s(T_{H,{\rm osc}})} 
    \frac{1}{N_A}
    R_M  \frac{R_F^2}{N_A} \Theta_{\Phi,\text{i}}^2,
    \label{eq: heavy abundance basis2} 
    \\
    \Omega_L 
    &\approx
    \frac{\chi_0}{2 \rho_{\rm crit}}
    \frac{s_0}{s(T_{L,{\rm osc}})}
    \frac{m_a(T_{{\rm osc}, L})}{m_{a,0}}
    R_M
    \left(\Theta_{A,\text{i}} - \frac{R_F^2}{N_A}\Theta_{\Phi, \text{i}}\right)^2 ,
    \label{eq: light abundance basis2}
\end{align}
where $m_{H,0}$ and $m_{L,0}$ are approximately equal to $m_{a,0}$ and $m_\phi$, respectively, if $r_f \gg 1$. Given that $R_F = r_f^{-1} \ll 1$, the heavy axion abundance looks much less than the light axion abundance. However, when $n_\phi = \mathcal{O}(1)$, we need to tune the parameter $N_A$ to be much smaller than unity. Based on the relation~\eqref{eq: f to F}, 
\begin{equation}
    n_\phi = \frac{\sin^2 \alpha_F}{N_A} \approx \frac{R_F^2}{N_A} = \mathcal{O}(1),
\end{equation}
we find that $N_A \ll 1$ is required, enhancing the heavy axion abundance. As a result, the same conclusion about the heavy axion abundance holds regardless of the choice of basis. However, fine-tuning  of the parameter $N_A \ll 1$ is necessary in the potential \eqref{eq: potential another}, which would require extra model-building efforts to implement in a natural way.

Conversely, we can consider the scenario in which the heavy axion abundance is suppressed, allowing the light axion to dominate dark matter~\cite{Kitajima:2014xla,Ho:2018qur}. For this scenario, the potential~\eqref{eq: potential another} is optimal; if $R_F \ll 1$ and $N_A = \mathcal{O}(1)$, $\Omega_H$ is much less than $\Omega_L$. In contrast, within the basis of Eq.~\eqref{eq: potential}, we have to set $n_\phi \ll 1$. It makes this potential less favorable. Thus, selecting the appropriate basis is essential when considering specific cosmological phenomena.

\section{Refined adiabatic condition}
\label{sec: refined adiabatic condition}

In this section, we examine the adiabatic condition in Eq.~\eqref{eq: adiabatic condition}.
We demonstrate that our definition of the level crossing is not only basis-independent but also provides a more rigorous formulation of the adiabatic condition. This improves upon some previous studies where the proposed adiabatic conditions were either basis-dependent or insufficient to fully characterize the adiabatic evolution.

First, we emphasize that the level-crossing condition~\eqref{eq: level crossing condition}, the definition of the level-crossing temperature~\eqref{eq: Tlc def}, and the timescale of the level crossing~\eqref{eq: time scale lc} are basis-independent. While the mixing angle changes under basis transformation (for example, in the $(A, \Phi)$ basis, it becomes $\alpha(T) + \alpha_F$), since $\alpha_F$ is temperature-independent, Eqs.~\eqref{eq: level crossing condition}, \eqref{eq: Tlc def}, and \eqref{eq: time scale lc} remain valid in any basis.

\begin{figure}[t!]
    \begin{center}  
        \vspace{5mm}
        \includegraphics[width = 121mm]
        {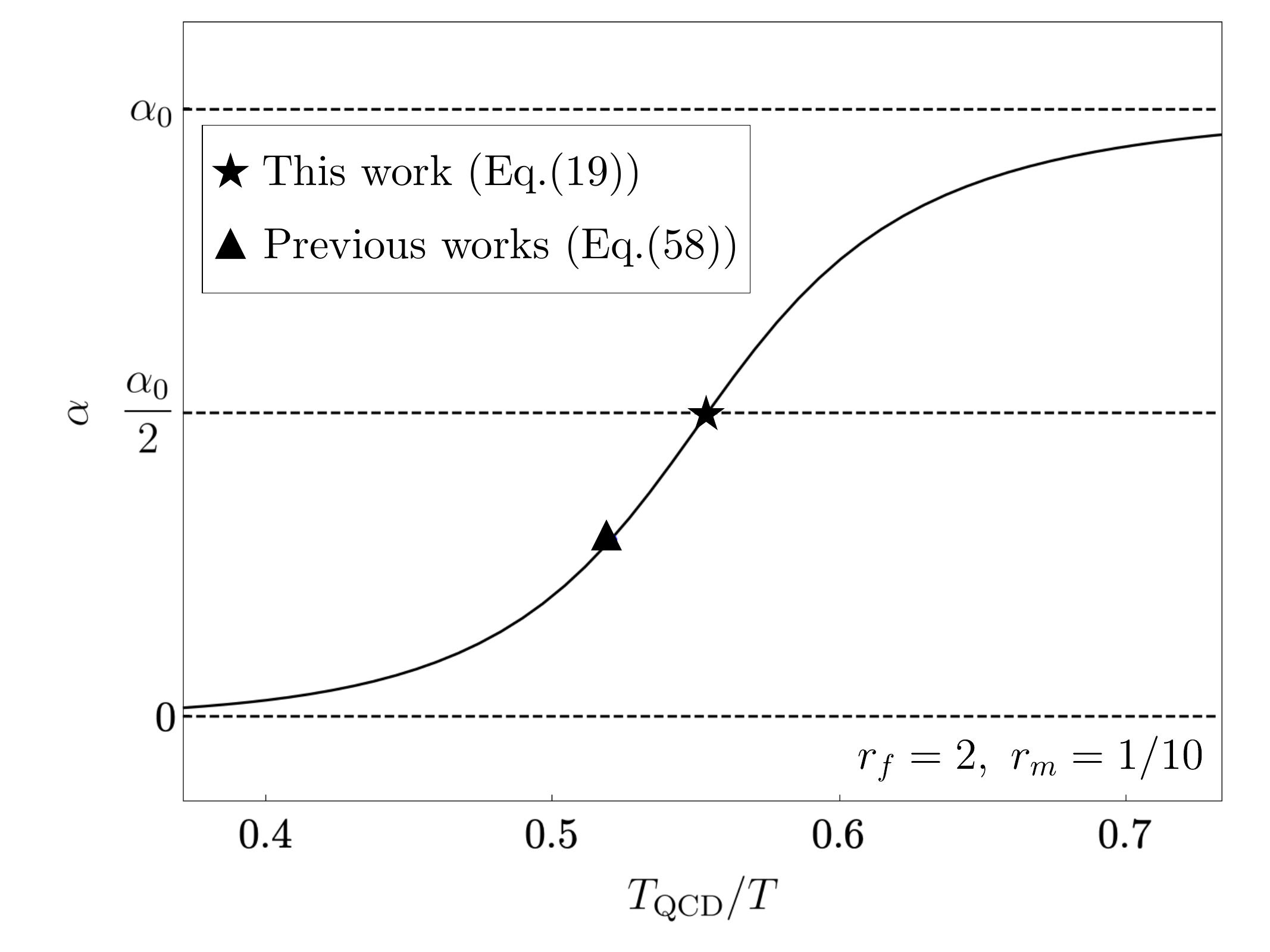}
    \end{center}
    \caption{%
    The temperature dependence of the mixing angle $\alpha(T)$ during the level crossing. The red point represents the mixing angle at the level-crossing temperature in Eq.~\eqref{eq: T level crossing}. It is obviously $\alpha_0 / 2$. The blue point represents the angle at the temperature defined by Eq.~\eqref{eq: T level crossing 2}~\cite{Ho:2018qur,Li:2023xkn,Li:2023uvt,Li:2024okl}. 
    }
    \label{fig: Tlc} 
\end{figure}

There are several studies on the level crossing and related phenomena~\cite{Kitajima:2014xla,Daido:2015bva,Daido:2015cba,Ho:2018qur,Murai:2023xjn,Cyncynates:2023esj,Li:2023xkn,Li:2023uvt,Li:2024psa,Li:2024okl}. Consequently, different definitions of $T_\times$ and $\Delta t_\times$ have been proposed in the literature. For example, some studies define the level crossing as the point where the difference of the mass eigenvalues squared reaches its minimum value~\cite{Ho:2018qur,Li:2023xkn,Li:2023uvt,Li:2024okl}, i.e., 
\begin{equation}
    \left. \frac{\text{d} (m_H^2 - m_L^2)}{\text{d} t} \right|_{T=T_\times} = 0.
    \label{eq: T level crossing 2}
\end{equation}
The mass eigenvalues are basis-independent, as they are physical quantities. Thus, this condition is also clearly basis-independent. However, if $r_f$ is around unity, the level-crossing temperature determined by this expression could deviate from the moment when the mixing angle changes most rapidly. We illustrate these two definitions of $T_\times$ and their relationship to the evolution of $\alpha$ in Fig.~\ref{fig: Tlc}. While our definition corresponds to the temperature when $\alpha$ varies most rapidly, at $T = T_\times$ given by Eq.~\eqref{eq: T level crossing 2}, $\alpha$ changes more slowly, resulting in a larger $\Delta t_\times$. Additionally, there exists parameter regions where $T_\times$ of Eq.~\eqref{eq: T level crossing 2} has no physical solution. Therefore, our definition \eqref{eq: Tlc def} is more robust and broadly applicable.

The timescale $\Delta t_\times$ has been characterized in various ways: as $H^{-1}(T=T_\times)$\cite{Kitajima:2014xla}, $|\text{d} \log{\cos{\alpha}} / \text{d}t|^{-1}_{T=T_\times}$\cite{Ho:2018qur}, and as the period during which off-diagonal terms are significant~\cite{Cyncynates:2023esj}. However, these expressions are either basis-dependent or fail to capture the true timescale of the level crossing.

The adiabatic condition for the level crossing varies across the literature.
In Ref.~\cite{Cyncynates:2023esj}, they use the adiabatic condition,
\begin{equation}
    \Delta t_\times \gg \frac{2 \pi}{m_L(T_\times)} \approx \frac{2 \pi}{m_\phi},
    \label{eq: adiabatic condition 1}
\end{equation}
 which requires that both axions oscillate many times during the level crossing. Here the second equality applies to the limit of $r_f \gg 1$. 
On the other hand, in Ref.~\cite{Ho:2018qur}, they consider an adiabatic condition that includes the beat frequency:
\begin{equation}
    \Delta t_\times \gg \max \left[ \frac{2 \pi}{m_L(T_\times)}, \frac{2 \pi}{m_H(T_\times) - m_L(T_\times)} \right] \approx \frac{2 \pi}{m_\phi} r_f,
    \label{eq: adiabatic condition 2}
\end{equation}
where the second equality applies to the case of $r_f \gg 1$. This condition requires that the time scale of the level crossing should be much longer than that of not only the axion oscillation but also the beat oscillation. It was also pointed out that while the total number $Y_H + Y_L$ is conserved, neither $Y_H$ nor $Y_L$ is individually conserved if $2 \pi / m_L(T_\times) < \Delta t_\times < 2 \pi / [m_H(T_\times) - m_L(T_\times)]$.

We numerically investigate which adiabatic condition better describes the level-crossing dynamics. Let us derive the upper limit on $f_a$ imposed by each adiabatic condition. We can rewrite the adiabatic conditions \eqref{eq: adiabatic condition 1} and \eqref{eq: adiabatic condition 2} in terms of $f_a$ as 
\begin{align}
    f_a & \lesssim \frac{1}{C_{\rm ad}} \frac{2 \sqrt{10 \chi_0}}{n \pi^2} \frac{M_{\rm pl}}{T_\mathrm{QCD}^2} \frac{1}{K(T_\times)} r_m^{1 + 4/n} r_f^{-1} \rm,
    \label{eq: fa adiabatic conditon 1} \\
    f_a & \lesssim \frac{1}{C_{\rm ad}} \frac{2 \sqrt{10 \chi_0}}{n \pi^2} \frac{M_{\rm pl}}{T_\mathrm{QCD}^2} \frac{1}{K(T_\times)} r_m^{1 + 4/n} r_f^{-2} \rm.
    \label{eq: fa adiabatic conditon 2} 
\end{align}
Here, $C_{\rm ad}$ is a dimensionless constant determined numerically based on the degree of adiabaticity we require.
While these upper bounds have the same dependence on $r_m$, they differ in their dependence on $r_f$.

\begin{figure}[t!]
    \begin{center}  
        \includegraphics[width = 81mm]
        {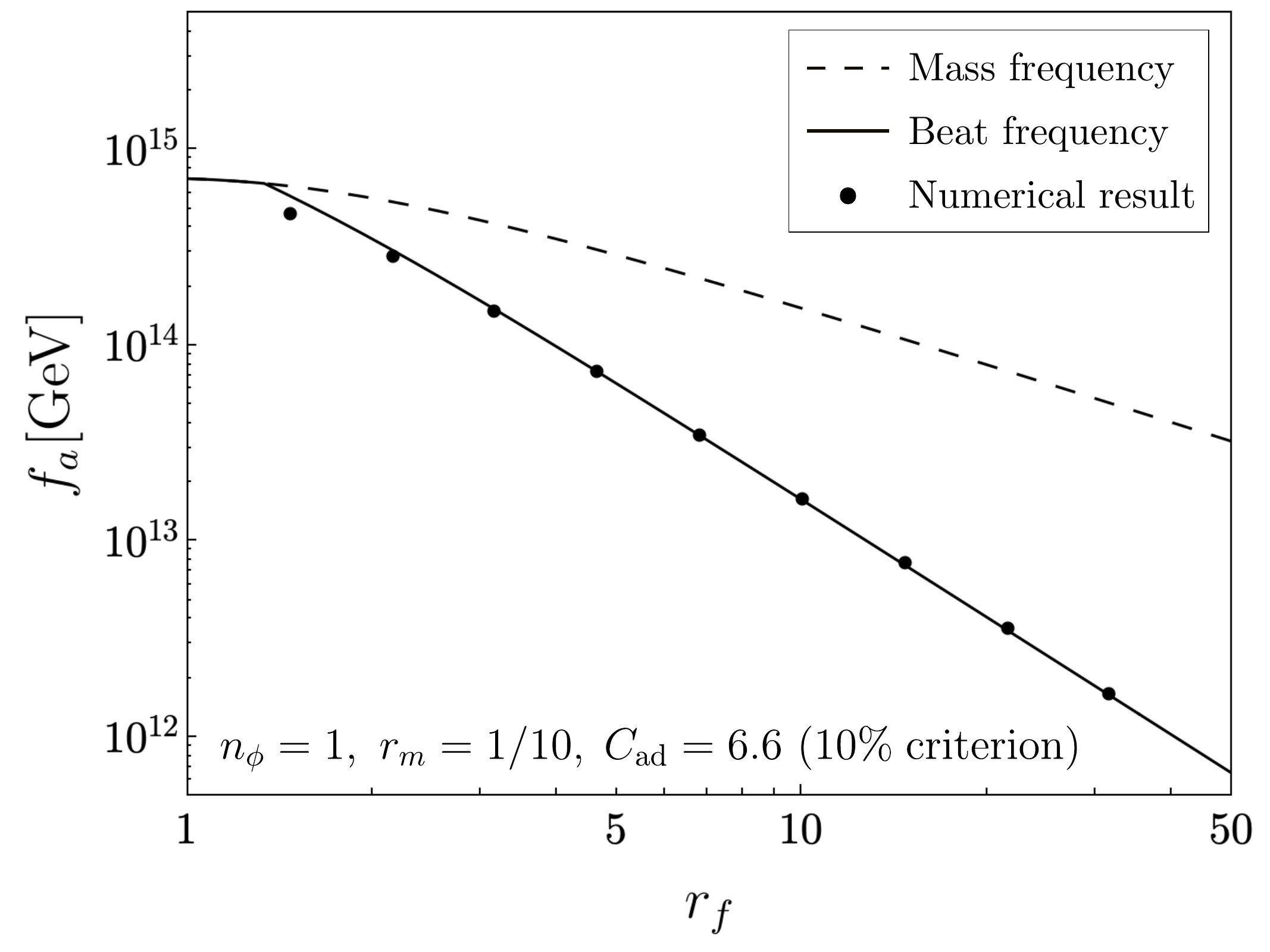}
        \includegraphics[width = 81mm]
        {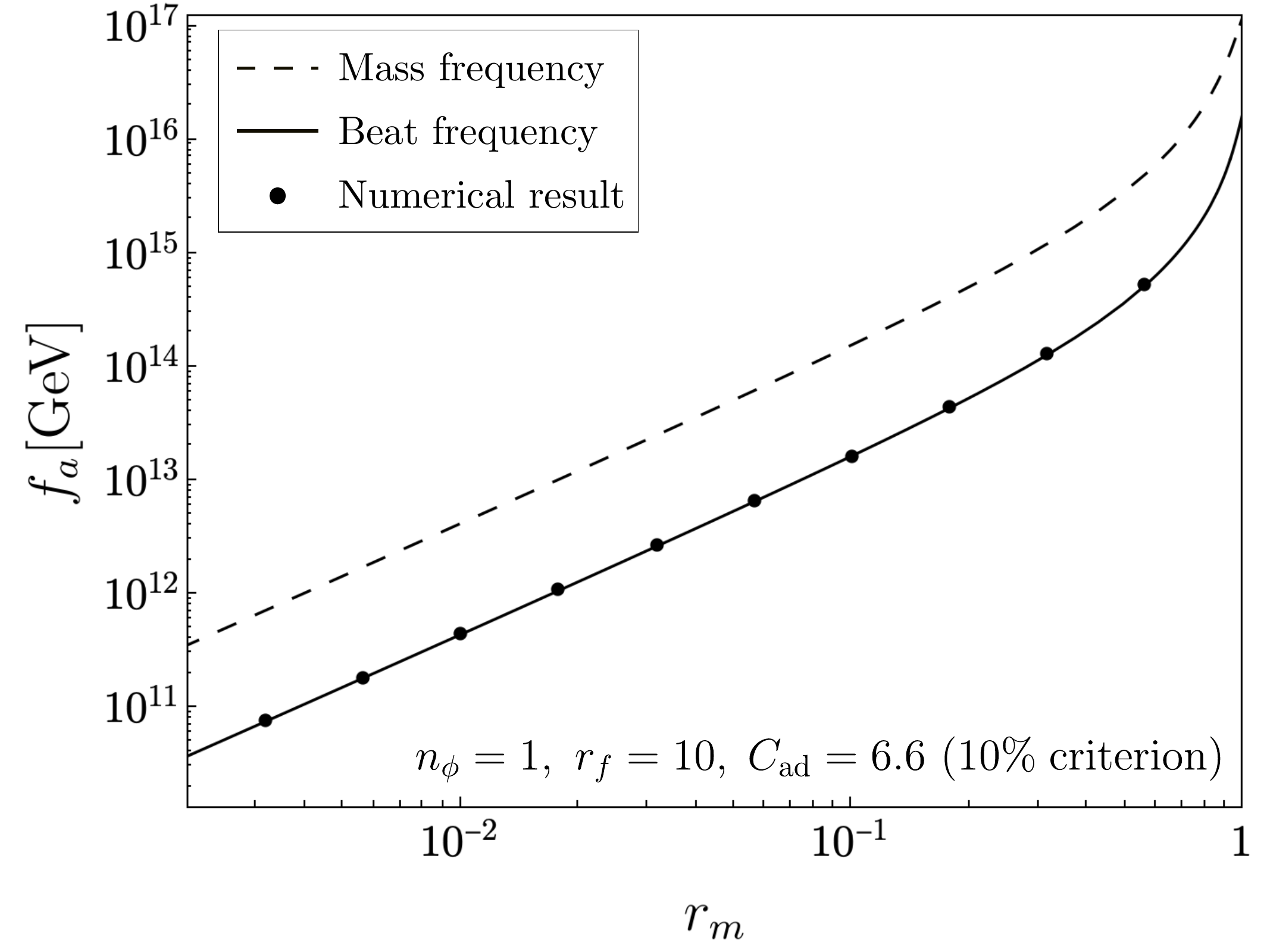}
    \end{center}
    \caption{The upper limits of $f_a$ in each adiabatic condition. The left panel shows the $r_f$-dependence for $r_m = 1/10$, and the right panel shows the $r_m$-dependence for $r_f = 10$. The numerical results (black points) rely on the $10 \%$ criterion. $C_{\rm ad}$ is determined for Eq.~\eqref{eq: fa adiabatic conditon 2} to fit the numerical results.
    }
    \label{fig: dependence} 
\end{figure}

To examine the dependence on $r_f$ and $r_m$, we numerically calculate the axion dynamics with the level crossing. Detailed numerical methods are described in Appendix~\ref{sec: numerical calculation}.
From these calculations, we obtain the energy densities of the heavy and light axions as functions of $T$, allowing us to calculate the number density transferred between the heavy and light fields. In Fig.~\ref{fig: dependence}, we present our numerical analysis of the adiabatic conditions. 
We set the initial conditions to $\theta_a = 1$ and $\theta_\phi = 0$, meaning that only the light mode is present in the initial time, and then evaluate the change in $Y_L$.
In particular, we derive the value of $f_a$ for fixed $r_m$ and $r_f$ with which $Y_L$ decreases by 10\% at the level crossing.
The numerical results show an $r_m$-dependence that is consistent with both analytical conditions. 
On the other hand, we find that the threshold value of $f_a$ depends on $r_f$ as $\propto r_f^{-2}$, which favors the condition~\eqref{eq: adiabatic condition 2}.
As a result, we demonstrated that the beat frequency plays a crucial role in the adiabatic condition.

\begin{figure}[t!]
    \begin{center}  
        \includegraphics[width = 80mm]
        {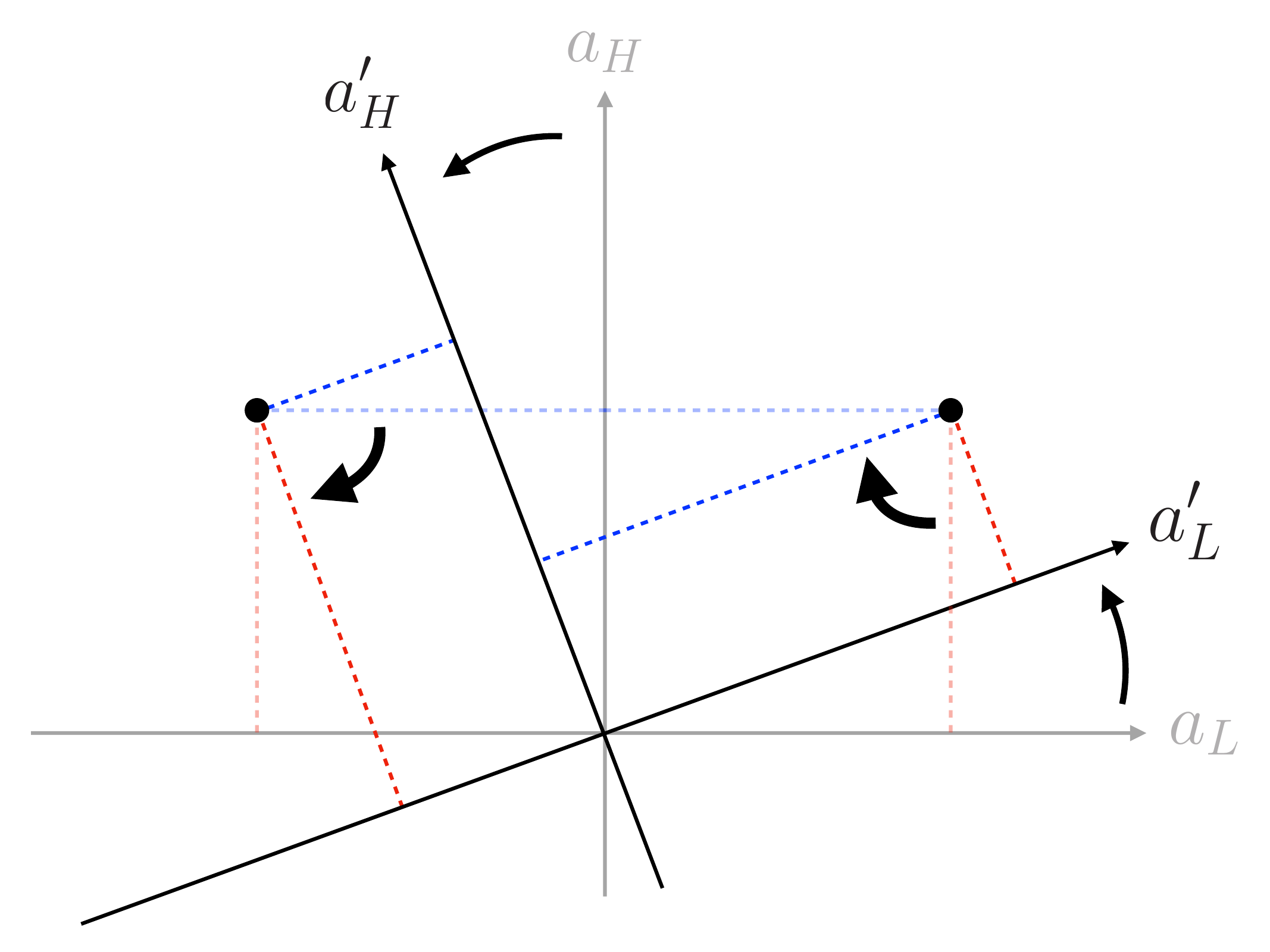}
    \end{center}
    \caption{The transfer of the number density between the heavy and light fields for non-adiabatic level crossing.}
    \label{fig: mass rotate} 
\end{figure}

The importance of the beat frequency can be qualitatively understood as follows. First, let us examine how the mass eigenstates transfer their number density during this process. To consider a situation where the adiabatic condition is not fully satisfied, let us consider an extreme case where the axions are nearly stationary during the evolution of $\alpha$. Fig.~\ref{fig: mass rotate} illustrates a rotation of the mass eigenstates at the level crossing. 
Note that the direction of the rotation follows the evolution of $\alpha(T)$ in Fig.~\ref{fig: level crossing}. The number density is directly related to the amplitude of the field oscillations. If the heavy and light fields have the same signs, the component of the light mode increases, and a part of $Y_H$ transfers to $Y_L$. 
Conversely, if the heavy and light fields have opposite signs, the opposite transfer occurs. As a result, the relative phase of the axion oscillations determines the transfer direction of the axion number.

\begin{figure}[t!]
    \begin{center}  
        \includegraphics[width = 110mm]
        {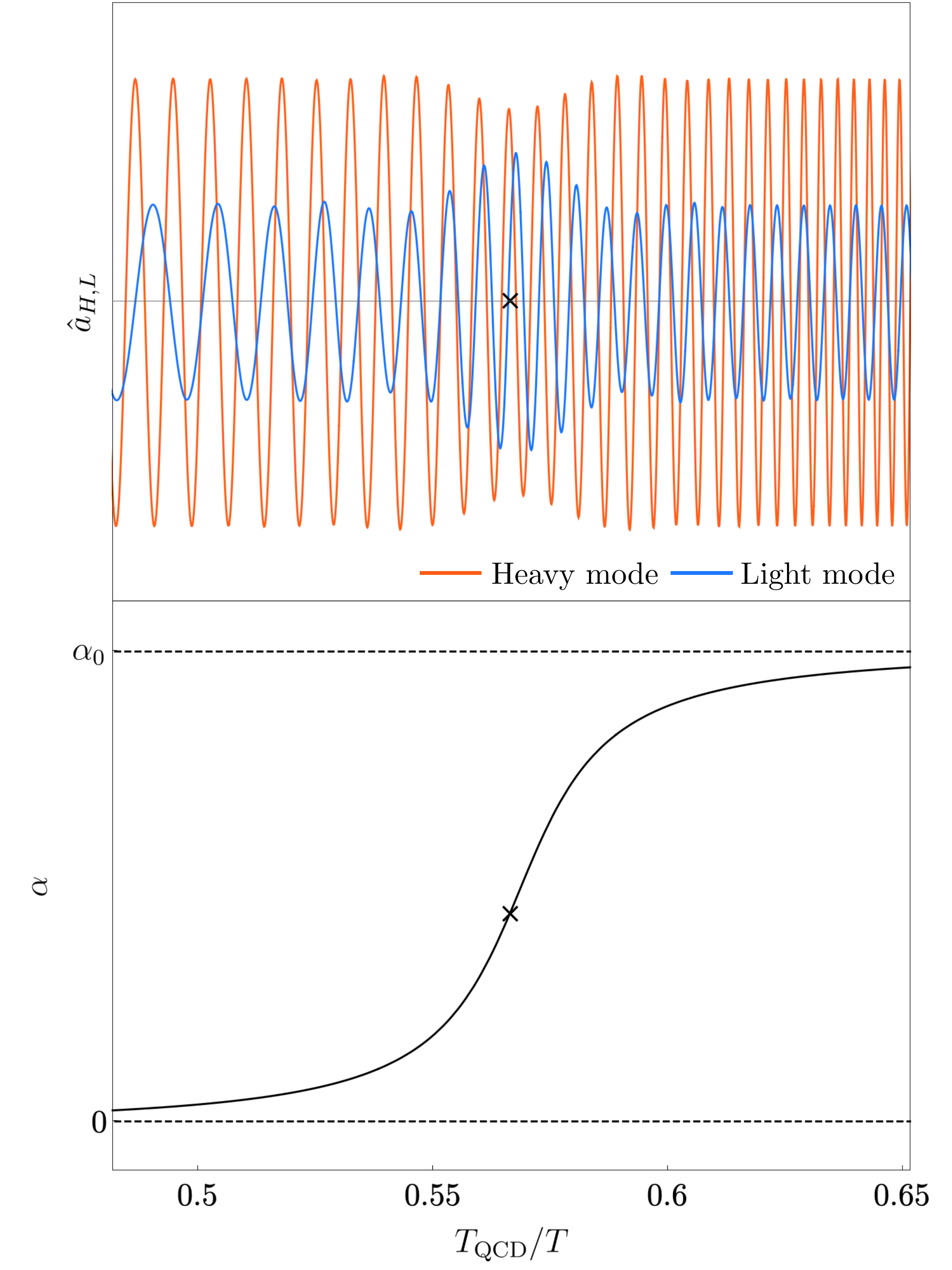}
    \end{center}
    \caption{
    The numerical result of the axion oscillations with $r_f = 10$, $r_m = 1 / 10$, $f_a = 1.8 \times 10^{13} \, \text{GeV}$, and initial conditions $\theta_a = \theta_\phi = 1$. The heavy and light modes are normalized as $\hat{a}_{H, L}(T) = \sqrt{m_{H, L}(T) / s(T)} \ a_{H, L}(T)$ to maintain constant amplitudes before and after the level crossing. The bottom panel shows the rotation of the eigenstates. The cross marks in both panels indicate $T_{\times}$.
    }
    \label{fig: numerical osc} 
\end{figure}

 The above discussion provides an intuitive understanding. In addition, the relation between the number transfer and the beat frequency is confirmed in a numerical calculation. We show an example of the axion dynamics during the level crossing in Fig.~\ref{fig: numerical osc}. Interestingly, we can adjust the input parameters, specifically the axion decay constant $f_a$, so that the total number transfer is negligibly small at the end, even if the adiabatic condition~\eqref{eq: adiabatic condition 2} is not satisfied. By tuning $f_a$, we can control the relative phase  between the two axions such that they oscillate in phase during the first part of the level crossing, while becoming out of phase afterwards. As shown in Fig.~\ref{fig: numerical osc}, where we consider a case where the adiabatic condition is marginally violated due to comparable time scales of beat frequency and level crossing, the number flows from the heavy mode to the light mode, which is clearly evident in the numerical results. Subsequently, due to the beat oscillation, the axion oscillations become in the opposite phase, leading to a reversal of the number transfer in the later part. This carefully controlled phase evolution ensures that the number transfers between light and heavy modes in the first and second halves of the crossing cancel each other out. In the figure, we specifically examine the case where, although the individual axion numbers are not conserved due to the marginal violation of the adiabatic condition, their sum remains conserved.
If multiple beats occur during the level crossing period, the system naturally tends toward adiabatic evolution as these repeated oscillations cause the number transfer to occur multiple times in alternating directions, effectively smoothing out the overall process.

\section{Discussion and Conclusions}
\label{sec: conclusion}
The QCD axion and ALPs are promising dark matter candidates, with the vacuum misalignment mechanism being one of their important production mechanisms. In single-axion models, the dark matter abundance is determined by the axion mass, decay constant, and initial misalignment angle. However, the presence of multiple axions can significantly modify this picture through their mixings.

We investigated the dynamics of two-axion systems, focusing on the level crossing that occurs in the early universe due to the temperature dependence of the QCD axion mass. Our analysis revealed that when the level crossing proceeds sufficiently slowly, a resonant conversion occurs between the two axions while conserving their comoving number densities. The importance of the beat frequency in the adiabatic condition for resonant axion conversion was first pointed out in Ref.~\cite{Ho:2018qur}. We significantly improved this understanding by numerically and analytically demonstrating why this condition, rather than the simpler conditions using axion mass scales~\cite{Cyncynates:2023esj,Kitajima:2014xla}, correctly describes the conversion dynamics. Furthermore, we developed basis-independent definitions for both the timing and duration of the level crossing, improving upon previous work~\cite{Ho:2018qur} where the definitions could give no solution in certain parameter regions. Our analysis provides a complete physical picture of why the beat frequency determines the adiabaticity of the level-crossing phenomenon.

Using this improved adiabatic condition and basis-independent analysis, we identified the viable parameter region where the QCD axion and ALP explain dark matter through the resonant conversion mechanism. An important feature of this scenario is that while the heavy axion's mass and photon coupling remain near the QCD axion band, those of the light axion can significantly deviate from it, providing distinctive experimental signatures.

Our study also clarified the relationship between different bases used to describe the level-crossing phenomenon. Although these bases are mathematically equivalent, we found that certain parameter choices that appear natural in one basis may require fine-tuning in another. This insight suggests that the choice of basis can be physically motivated when considering specific scenarios, such as the enhancement of the QCD axion abundance.

For simplicity, we assumed a time-independent ALP potential in our analysis. Generalizing to time-dependent potentials, as suggested in Refs.~\cite{Arias:2012az,Nakagawa:2022wwm}, could delay the onset of oscillations and enhance the ALP abundance. In such cases, a smaller ALP decay constant than shown in Figs.~\ref{fig: reliable region} and \ref{fig: reliable region2} could explain the observed dark matter abundance, with the values presented in our analysis representing the upper limits. Intriguingly, we identified parameter regions where both axions contribute significantly to dark matter with similar decay constants, making them simultaneously accessible to experimental probes.

\section*{Acknowledgments}
FT thanks David Cyncynates and Jed Thompson for useful communications regarding the choice of basis.
This work is supported by JSPS Core-to-Core Program (grant number: JPJSCCA20200002) (F.T.), JSPS KAKENHI Grant Numbers 20H01894 (F.T.), 20H05851 (F.T. and W.Y.), 21K20364 (W.Y.), 22H01215 (W.Y.), 22K14029 (W.Y.), 23KJ0088 (K.M.), and 24K17039 (K.M.), Graduate Program on Physics for the Universe (Y.N.), JST SPRING, Grant Number JPMJSP2114 (Y.N.), Incentive Research Fund for Young Researchers from Tokyo Metropolitan University (W.Y.). 
This article is based upon work from COST Action COSMIC WISPers CA21106, supported by COST (European Cooperation in Science and Technology).

\appendix 

\section{Numerical calculation}
\label{sec: numerical calculation}

Here, we summarize the equations of motion for two axions used to numerically evaluate the adiabaticity of the level crossing. The time evolution of the QCD axion $a$ and the ALP $\phi$ is described by the equations of motion derived from the potential $V(a, \phi)$ in Eq.~\eqref{eq: potential},
\begin{align}
    \frac{\text{d}^2 a}{\text{d} t^2} + 3 H (T) \frac{\text{d} a}{\text{d} t} + \frac{\partial V(a, \phi)}{\partial a} &= 0,
    \\
    \frac{\text{d}^2 \phi}{\text{d} t^2} + 3 H (T) \frac{\text{d} \phi}{\text{d} t} + \frac{\partial V(a, \phi)}{\partial \phi} &= 0,
\end{align}
where the Hubble parameter $H(T)$ is given by the Friedmann equation \eqref{eq: Friedmann eq}. To reduce computational cost, we introduce the normalized inverse temperature, $\tau \equiv T_\text{QCD} / T$. We treat this parameter as a substitute for the cosmic time $t$, using the relation of
\begin{equation}
    \frac{\text{d} \tau}{\text{d} t} = \frac{\pi}{\sqrt{10}} \frac{T_\text{QCD}^2}{M_\text{pl}} \tilde{K}(\tau),
    \label{eq: t to tau}
\end{equation}
where $\tilde{K}(\tau)$ is a function of $\tau$ introduced to make the equations simpler,
\begin{equation}
    \tilde{K}(\tau) \equiv \frac{\sqrt{g_*(\tau)}}{\tau} \frac{g_{*s}(\tau)}{3 g_{*s}(\tau) - \tau \dfrac{\text{d} g_{*s}(\tau)}{\text{d} \tau}}.
    \label{eq: tilde K}
\end{equation}
Then, the equations of motion become
\begin{align}
    \tilde{K}(\tau) \frac{\text{d}^2 \theta_a}{\text{d} \tau^2} + \tilde{K}(\tau) \left(\frac{\text{d} \tilde{K}(\tau)}{\text{d} \tau} + \sqrt{g_*(\tau)} \tau^{-2} \right)\frac{\text{d} \theta_a}{\text{d} \tau} + \frac{10 M_\text{pl}^2}{\pi^2 T_\text{QCD}^4 f_a^2} \frac{\partial V(\theta_a, \theta_\phi)}{\partial \theta_a} &= 0,
    \\
    \tilde{K}(\tau) \frac{\text{d}^2 \theta_\phi}{\text{d} \tau^2} + \tilde{K}(\tau) \left(\frac{\text{d} \tilde{K}(\tau)}{\text{d} \tau} + \sqrt{g_*(\tau)} \tau^{-2} \right)\frac{\text{d} \theta_\phi}{\text{d} \tau} + \frac{10 M_\text{pl}^2}{\pi^2 T_\text{QCD}^4 f_\phi^2} \frac{\partial V(\theta_a, \theta_\phi)}{\partial \theta_\phi} &= 0.
\end{align}
Here, we define $\theta_a = a /f_a$ and $\theta_\phi = \phi / f_\phi$. We refer to Ref.~\cite{Saikawa:2018rcs} for the detailed temperature dependence of $g_*$ and $g_{*s}$. In our calculations, we take the initial field values as $\theta_\phi(\tau_\text{i})=1$, $\theta_a(\tau_\text{i})=0$, where $\tau = \tau_i$ represents the inverse temperature well before the ALP begins to oscillate. 
To precisely evaluate the adiabaticity of the level crossing, we set one of the initial modes to zero. This is because slight variations in parameters can significantly impact energy transfer between each mass eigenstates when the level crossing is non-adiabatic. By solving these equations, we obtain the time evolution of $\theta_\phi(\tau)$ and $\theta_a(\tau)$ during the level crossing, which determines the number densities, $N_{H/L}(\tau)$, as a function of $\tau$. To numerically estimate the adiabaticity, we compare the axion yield parameters before and after the level crossing, using the following parameter,
\begin{equation}
    R_\text{ad} \equiv \frac{|Y_H(\tau_a) - Y_H(\tau_b)|}{Y_H(\tau_b) + Y_L(\tau_b)} = \frac{|Y_H(\tau_a) - Y_H(\tau_b)|}{Y_H(\tau_a) + Y_L(\tau_a)},
\end{equation}
where $\tau_b$ and $\tau_a$ denote the inverse temperature at times just before and after the level crossing, respectively. Now, we take $\tau_a = \tau_\times + \frac{1}{2} \Delta \tau_\times$ and $\tau_b = \tau_\times - \frac{1}{2} \Delta \tau_\times$, where $\tau_\times$ corresponds to the parameter $\tau$ at the level-crossing temperature from Eq.~\eqref{eq: T level crossing} and $\Delta \tau_\times$ represents the timescale of the level crossing which is given by Eq.~\eqref{eq: Delta t cross} and Eq.~\eqref{eq: t to tau}.
We note that the total number $Y_H + Y_L$ remains conserved despite the level crossing as both axions have already begun oscillating beforehand. In this paper, we define the threshold for the adiabaticity breaking as $R_\text{ad} = 0.1$, applying a 10\% criterion.

\section{Friction in chiral perturbation theory}
\label{sec:friction}
\label{app:friction}

To discuss the friction on the axion, whose mass is much smaller than the cosmic temperature, in the chiral perturbation theory,
we apply {the}
method originally discussed in the context of warm inflation with renormalizable interaction~\cite{Berera:1995ie,Berera:1998gx,Yokoyama:1998ju,Bastero-Gil:2012akf}. This method is also applied to the oscillating scalar field in a thermal environment relevant to dark matter production~\cite{Nakayama:2021avl}. 
The friction {is} 
interpreted as a thermalization {process} of the ambient plasma, whose dispersion relation depends on the slowly moving scalar field, driving the system toward a time-dependent equilibrium state. 

Here we consider the potential in the following well-known form (see e.g.~\cite{GrillidiCortona:2015jxo,Takahashi:2023vhv})
\begin{align}
\label{inst}
    V^{(A\pi)}(A,\pi)= -m_\pi^2 f_\pi^2 \sqrt{1-\frac{4m_um_d}{(m_u+m_d)^2}\sin^2 \left( \frac{A}{2f_A} \right)}  \cos \left[ \frac{\pi}{f_\pi} - \phi_A \right],
\end{align}
with
\begin{align}
    \tan \phi_A \equiv \frac{m_u-m_d}{m_d+m_u} \sin \left[ \frac{ A}{2 f_A} \right],
\end{align}
in the two flavor case. Here $\pi$ is the neutral pion, $m_\pi \approx 135\,\rm MeV$ is the neutral pion mass, and $f_\pi\approx 93\,\rm MeV$ is the pion decay constant. $m_u$ and $m_d$ are the masses of the up and down type quarks, respectively.
By integrating out the pion, we would get the standard axion potential. {Taking the leading contribution of the cosine function, we obtain the zero temperature limit of the first term of Eq.~\eqref{eq: potential}. Here, we do integrate out the pion.}

By treating $A$ as background, the pion axion potential  is approximated as follows:
\begin{align}
    {V(A, \vec{\pi})}
    \simeq \left( -f_\pi^2+\frac{\vec{\pi}^2}{2} \right) m_\pi^2 \sqrt{1-\frac{4m_um_d}{(m_u+m_d)^2}\sin^2 \left( \frac{A}{2f_A} \right)} ,
\end{align}
where we redefined the neutral pion as $\pi\to \pi-f_\pi \phi_A$.
The mass term is not only for neutral pion but also for the charged pions due to the isospin symmetry (see, e.g., Ref.~\cite{Ubaldi:2008nf}).

Using the potential, the equation of motion for the axion is given as 
\begin{align}
    \ddot A+3 H \dot A= -\partial_A V(A, \vec \pi)   = \left( f_\pi^2-\frac{\vec{\pi}^2}{2} \right)\partial_A m_\pi^2 (A) .
\end{align}
In thermal environment, we can take the thermal average of $\vec\pi^2$ in the right-handed side
\begin{align}
    \langle\vec \pi^2 \rangle= 3 \int \frac{d^3 \vec p}{(2\pi)^3  \sqrt{\vec p^2+m_\pi^2(A) }} f(\vec p) ,
\end{align}
where we defined the axion background dependent pion mass squared 
\begin{align}
    m_{\pi}^2(A)\equiv m_\pi^2 \sqrt{1-\frac{4m_um_d}{(m_u+m_d)^2}\sin^2 \left( \frac{A}{2f_A} \right)}.
\end{align}

If the axion were not moving, we could take the distribution function $f$ by the thermal one, i.e., the Bose-Einstein distribution, $f_\mathrm{eq}$. However, in the oscillating background, it is not completely true, i.e.,
$f\neq f_{\rm eq}$ due to the motion of $A$. The pion distribution $f$ with $A$ is given by the pion distribution slightly before due to the finite timescale of thermalization $\Delta t_{\rm th}^{(\pi)}\sim (\frac{T^3}{4\pi^3 f_\pi^2})^{-1}$ for pions:
\begin{align}
    f\simeq f_{\rm eq}- \dot{A} \frac{\partial f_{\rm eq}}{\partial A} \Delta t_{\rm th}^{\rm (\pi)}.
\end{align}
Then approximating the distribution by the Boltzmann distribution and expanding by $m_\pi^2$ assuming $m_\pi< T$, we can get the analytic formula for $\langle\vec \pi^2 \rangle$:
\begin{align}
\langle\vec \pi^2 \rangle=\langle\vec \pi^2 \rangle_{\rm eq} +\frac{3 }{4\pi^2} \dot A \Delta t_{\rm th}^{(\pi)} \partial_A m_{\pi}^2.
\end{align}

Therefore the equation of motion for the axion can be obtained with the additional friction term 
\begin{align}
    \ddot A+(3 H +\Gamma_{\rm dis})\dot A= -\partial_A V_{\rm QCD},
\end{align}
with 
\begin{align}
    \Gamma_{\rm dis}= \Delta t_{\rm th}^{{(\pi)}} \frac{3(\partial_{A} m_\pi^2(A))^2 }{8\pi^2}\sim \frac{3\pi f_\pi^2(\partial_{A} m_\pi^2(A))^2 }{2 T^3}.
\end{align}
Here we consider that $V_{\rm QCD}$ from the (semi-)lattice QCD simulation that we used in the main text
involves the terms of the thermal equilibrium contribution of pion, $\langle \vec{\pi}^2\rangle_{\rm eq}$, and the vacuum contribution. 

\begin{figure}[t!]
    \begin{center}  
        \includegraphics[width = 110mm]
        {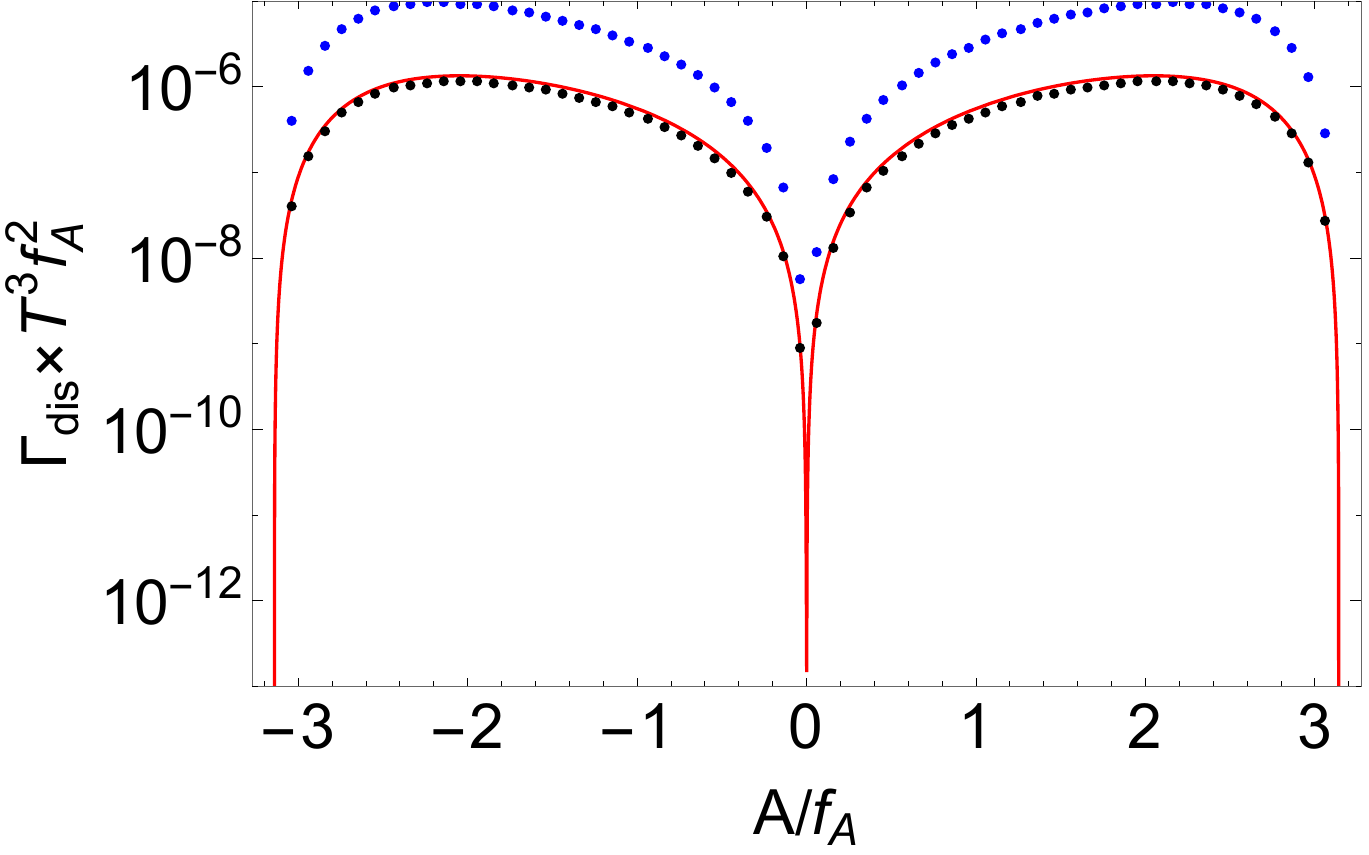}
        \includegraphics[width = 110mm]
        {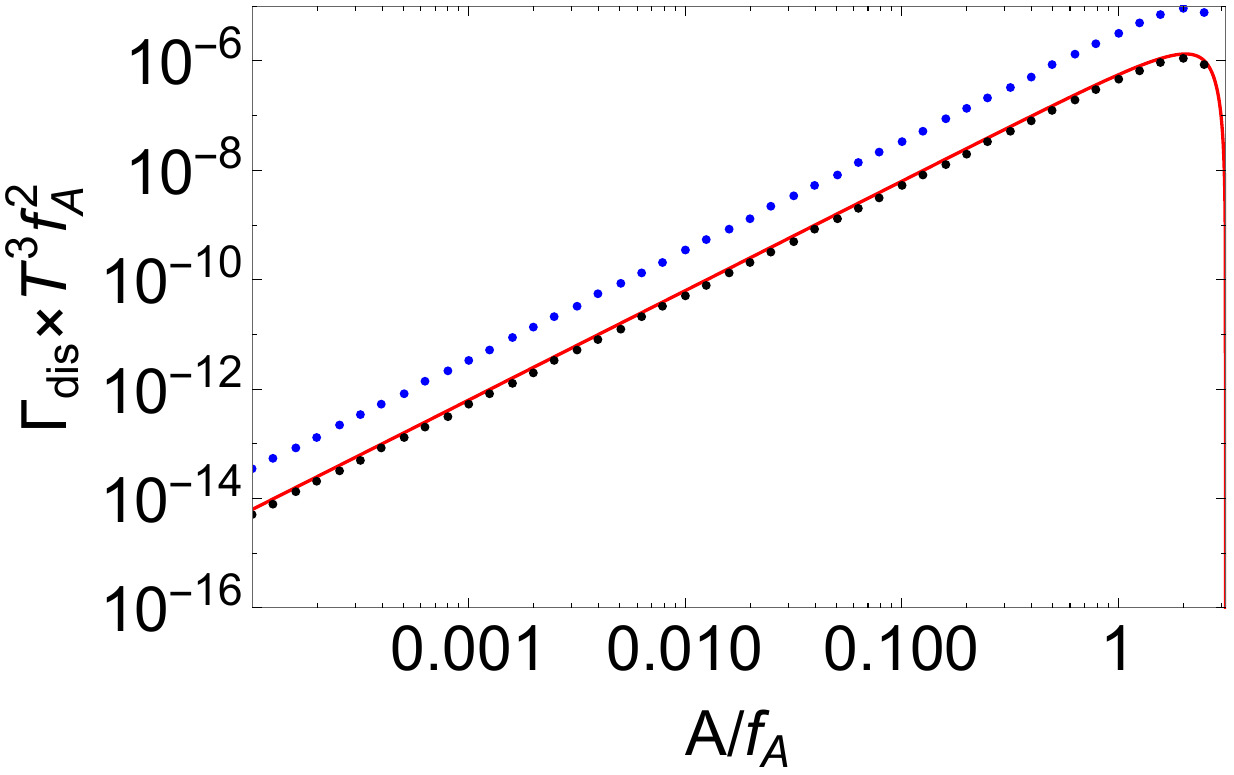}
    \end{center}
    \caption{The dissipation rate ($\times T^3 {f_A^2}$ 
    in GeV$^{6}$) by varying $A/f_A$. 
    The red solid line denotes the analytic formula we derived. The black (blue) points are the numerical evaluation without taking the leading $m_\pi^2$ expansion and we take $T=1\,\rm GeV$ with using the Boltzmann (Bose-Einstein) distribution for $f_{\rm eq}$. 
    The upper and lower panels are the same but with different scaling in $A/f_A$. 
    We used $m_u = 2.16\, {\rm MeV}$,  $m_d = 4.70\, {\rm MeV}$, $m_\pi^0 = 135\,{\rm MeV}$, and $f_\pi = 93 \, {\rm MeV}$.
}  \label{fig:dis} 
\end{figure}

\begin{figure}[t!]
    \begin{center}  
        \includegraphics[width = 110mm]
        {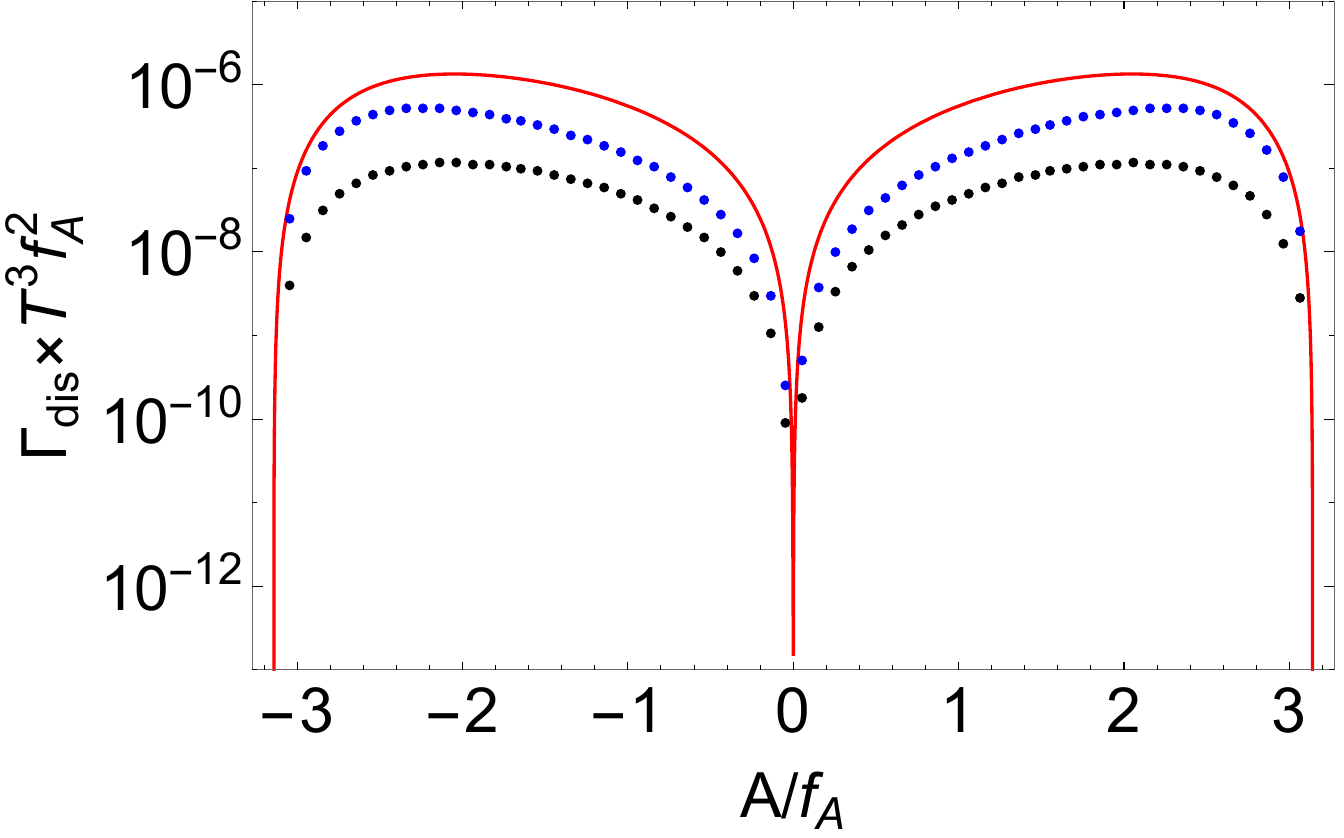}
        \includegraphics[width = 110mm]
        {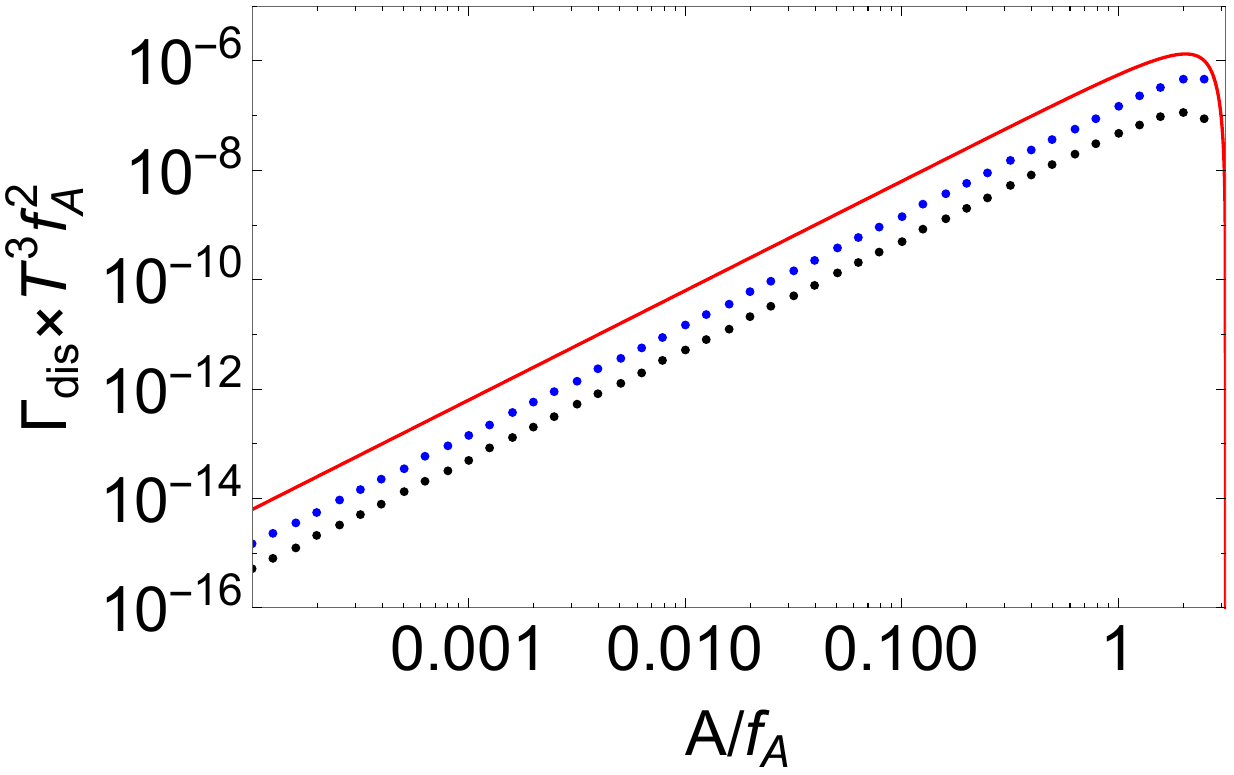}
    \end{center}
    \caption{Same as Fig.~\ref{fig:dis}, but $T=100$\,MeV for the numerical simulations.
}  \label{fig:dis2} 
\end{figure}

In Figs.~\ref{fig:dis} and \ref{fig:dis2}, we plot the dissipation rate that we derived by varying $A$. The analytic result agrees well with the ones performing integral numerically. Removing the approximation of the Boltzmann distribution can enhance the rate by an order of magnitude. 
We also emphasize that when the axion field value is not small this contribution can be significant. 

Lastly we comment on a possible thermal effect that we did not take into account. With a high temperature, the pion has thermal reactions with, e.g., nucleons or photons. This is usually not diagonal in the mass of the vacuum. 
Indeed, in the basis we have chosen in \eqref{inst}, the derivative coupling of axion in the kinetic term is isospin singlet~\cite{GrillidiCortona:2015jxo}.
In this case, $\pi$ or rather than  $\pi-\phi_Af_\pi$ is produced from the nucleon interaction or pion self-interaction from the derivative terms involving the kinetic term. Since the scattering time scale, $\frac{1}{4\pi^3}(1/f_\pi)^2 T^3, {4\pi^3}(m_N/f_\pi)^2 T$, can be faster than the mass scale $m_\pi$ at $T\gtrsim m_\pi$, the thermal plasma is dominantly composed of the $\pi$ particles rather than $\pi-\phi_A f_\pi$. In other words, it is not completely the mass eigenstate, $\pi$, in the high-temperature regime.  
From the cosine-term $\cos (\pi/f_\pi-\phi_A)$, we can obtain a four-point interaction involving a single axion, which provides the thermal dissipation contribution.

Then we estimate the imaginary part of the two-point function of axions, which is represented by the sunset diagram with thermal correction. Because the Briet-Wigner form of the propagator of pion is 
\begin{align}
\Delta_\pi \sim \frac{i}{q^2-m_\pi^2+i m_\pi \Gamma_{\rm th}},
\end{align}
with 
$\Gamma_{\rm th}$ being the thermalization rate of the pion. 
Precisely estimating the contributions beyond the scope of the paper, which is a definitely interesting topic and will be performed elsewhere, but here we assume $T\gtrsim m_\pi$ and $m_\pi\lesssim \Gamma_{\rm th}$ to do the estimation very na\"{i}vely. 

Then the imaginary part of the sunset diagram does not vanish even if the external momenta of axions are extremely small due to the large imaginary part of the propagator. By using $f_\pi\sim m_\pi$,
we get 
\begin{align}
\Gamma_{\rm dis,2}\sim \left(\frac{m_\pi}{f_a}\right)^2 T,
\end{align}
which may even not depend on the axion abundance. 

\section{Axion mixing and heavy and light QCD axion}
\label{app:mix}
In this section, we analyze the mixing effects between the QCD axion and ALP in vacuum, which is distinct from the dynamical phenomena like level crossing or resonant conversion discussed in the main text. We show how this mixing leads to the emergence of a heavy {or light} QCD axion and discuss its experimental implications.

In the vacuum, the second term of Eq.~\eqref{eq: potential another} generates a mass mixing between the QCD axion and ALP:
\begin{align} V_{\rm mix} \simeq M_\Phi^2 F_\Phi^2 N_A \frac{A}{F_A} \frac{\Phi}{F_\Phi}\equiv M_{\rm mix}^2 \Phi A .
\end{align}
{Here we assume $M_A^2 \gg M_\Phi^2, M_{\rm mix}^2$.} Then,
the mixing angle is approximately given by 
\begin{align}
\theta_{\rm mix}\simeq  \frac{M_{\rm mix}^2}{M_A^2}= N_A
\frac{\Lambda^4 }{\chi} \frac{F_A}{F_\Phi},
\end{align}
where we defined the dynamical scale $\Lambda^4= M_\Phi^2 F_\Phi^2$.
The QCD axion $A$ couples to various Standard Model particles $i$ with coupling strength $g_{Ai}=c_i/F_A$. Through the mixing effect, the $\Phi$ field inherits these couplings with strength
\begin{align}
g_{\Phi i}= N_A\frac{\Lambda^4}{\chi}\frac{c_i}{F_\Phi}
= \frac{c_i}{F_{\Phi {\rm eff}}},
\end{align}
where we have defined the effective decay constant $F_{\Phi\rm eff}= \frac{\chi}{N_A \Lambda^4}F_\Phi$.
Due to its gluon coupling, this mixed state represents a heavier {or lighter} variant of the QCD axion. {This can be seen by calculating the product of the mass and the effective decay constant},
\begin{align}
    m_\Phi F_{\Phi\rm eff} &= \frac{1}{N_A} \left(\frac{\sqrt{\chi}}{\Lambda^2}\right) \sqrt{\chi}.
\end{align}
Thus, for $N_A = {\cal O}(1)$, the $\Phi$ field becomes the heavy (light) QCD axion if $\chi > \Lambda^4~(\chi < \Lambda^4)$.
Due to the assumption on the axion masses, $\Phi$ becomes the heavy QCD axion if $F_A > F_\Phi$. On the other hand, the light QCD axion requires $F_A \ll F_\Phi$.

Importantly, this heavy or light axion couples to the nucleon electric dipole moment in the same way as the conventional QCD axion. Therefore, experiments searching for axion-induced oscillating effects may detect this variant QCD axion dark matter rather than the conventional QCD axion.

\bibliographystyle{apsrev4-1}
\bibliography{ref}

\end{document}